\documentclass[useAMS,usenatbib,usegraphicx]{mn2e}
\usepackage[british]{babel}
\usepackage{times}
\usepackage{psfig,epsfig}
\newcounter{subfigure}

\title[The nature of late-type spiral galaxies: structural parameters,
colour profiles, and dust extinction]
{The nature of late-type spiral galaxies: structural parameters, 
optical and near-infrared colour profiles, and dust extinction}

\author[Ganda et al.]{Katia Ganda $^1$,
Reynier F.\ Peletier$^1$\thanks{E-mail: peletier@astro.rug.nl},
Marc Balcells$^2$ and Jes\'us Falc\'on-Barroso$^{3}$\\
$^1$Kapteyn Astronomical Institute, Postbus 800, 9700 AV Groningen,The Netherlands\\
$^2$Instituto de Astrof\'\i sica de Canarias, Via Lactea s/n, 38700 La Laguna, Tenerife,Spain\\
$^3$European Space Agency / ESTEC, Keplerlaan 1, 2200 AG Noordwijk, The Netherlands}

\date{Released 2009 Xxxxx XX}

\pagerange{\pageref{firstpage}--\pageref{lastpage}} \pubyear{2009}

\def\LaTeX{L\kern-.36em\raise.3ex\hbox{a}\kern-.15em
    T\kern-.1667em\lower.7ex\hbox{E}\kern-.125emX}

\begin{document}

\label{firstpage}

\maketitle

%\mbremark{have added suggested new words in title}

\begin{abstract}
We analyse $V$ and $H$-band
surface photometry of a sample of 18 Sb-Sd galaxies. Combining high resolution HST images with
ground-based NIR observations, we extract photometric profiles, which cover the whole disk and
provide the highest possible resolution. This is the first photometric study of
late-type spirals for which the stellar kinematics have been measured. For 10 out of the 18
galaxies, HST data in both F160W ($H$) and F606W ($V$) are available, and, for those, we present colour maps and radial colour profiles at the resolution of the Hubble Space Telescope.

%We show that there is a considerable range in colours from galaxy to galaxy.
%The individual objects present in several cases some jumps in the inner regions, 
%but are otherwise rather flat. 
Colours vary significantly from galaxy to galaxy, but tend to be highly homogeneous within each
galaxy, with smooth and flat colour profiles.   Some of the colour maps show jumps in the inner
regions, likely due to dust.   We determine extinction-maps in an almost model-independent way
using the $V-H$ colour map and the SAURON Mg~b absorption line map of Ganda et al. (2007). The
maps show that A$_V$ ranges from 0 to 2 mag, in the center from 0 to 1.5 mag, in agreement with
the models of Tuffs et al. (2004).

We describe the surface brightness profiles as the superposition of an exponential disk and
a S\'ersic bulge.  The bulges are small (0.1-2.5 kpc), and show a
%and find that the bulges are in many cases very small, with the
shape parameter $n$ ranging from $\approx$ 0.7 to 3, with a mean value
smaller than two: well below the value for the {\sl classical} de Vaucouleurs
bulges. Most galaxies (16 out of 18) show a central light excess above the S\'ersic fit
to the bulge, which can be interpreted as a
nuclear cluster, as shown by previous studies. We provide zero-order
estimates for the magnitude of these components. 
We discuss the correlations among the structural galaxy parameters and with
other relevant quantities  such as Hubble type and stellar velocity dispersion.
We compare these results  with a recent paper by Graham \& Worley (2008), who
present a summary of most of the near-IR surface photometry of spirals in the
literature. 
%We find a strong correlation between central velocity dispersion and bulge luminosity, 
For both early and late type spirals, 
bulge luminosity strongly correlates with central velocity dispersion;   
at constant velocity dispersion, later-type bulges 
%The difference between early
%and late-type bulges of the same velocity dispersion is that late type bulges
are larger and less dense, and have lower S\'ersic $n$ values.
%Our data seem to support a scenario where the structural properties
%of bulges and discs are related to the mass of the bulge and where the shape
%parameter $n$ increases from $\approx$ 1 (exponential profile) to $\approx$ 4
%(de Vaucouleurs profile) as the galaxy age and the bulge grow. 
\end{abstract}


\section[Introduction]{INTRODUCTION}
It has been known for many years that the inner regions of spiral galaxies are
fundamentally different from their outer parts. The inner parts are generally
rounder, redder, and show a higher fraction of random vs. ordered motion.
Images of spirals show that less star formation  is present  in the central regions. Radial
surface brightness profiles can be fitted well by a large exponential disk and a
central, steeper component (e.g., Kent 1984). When looking at the stellar 
kinematics, one finds that v/$\sigma$ in the inner parts is much lower than in
the outer parts, where the rotation velocity is high and the stellar dispersion
low (e.g., Noordermeer 2006). As a result, people 
traditionally viewed spiral galaxies as a combination of a flattened disk
and a spheroidal central component called `bulge', which are assumed to be
physically and dynamically different: in this picture, the disk
component is rotationally supported against gravity, and the bulge is a hotter
system, similar to an elliptical galaxy. To understand how spiral galaxies are
built, one generally does a bulge-disk decomposition and studies the parameters
of both components. If one wants to go into more details, one can go one step
further, and also study  possible components
such as bars, spiral arms, rings,  and inner disks (see for example 
\citealt{jong96bis}, \citealt{prieto} and Erwin \& Sparke 2002).

When doing the bulge-disk decomposition, it is imperative that the decomposition 
is well-defined. The bulge-disk decomposition can be based e.g. on the surface 
brighteness profile
(photometric decomposition), the stellar kinematics (kinematic decomposition) or
on the two-dimensional axis ratio (morphological definition) (see Peletier
2008). The physical interpretation of the resulting parameters then depends
strongly on the definition. Up to now most B/D decompositions have used the
photometric decomposition, using a one-dimensional azimuthally averaged surface
brightness profile. Such profiles can generally be fitted well by an exponential
disk and a central S\'ersic r$^{1/n}$ distribution (Andredakis, Peletier \& Balcells 1995,
Graham 2001). The S\'ersic index $n$ has been found to correlate with the
morphological type of the galaxy, and generally ranges between 1 and 4. For a 
comprehensive review about the S\'ersic index see Graham \& Driver (2005). In
recent years it has been discovered that the bulges with surface brightness
profiles that are close to exponential (i.e. S\'ersic index close to 1) are
different from bulges with larger S\'ersic indices, which resemble more
elliptical galaxies. Bulges with exponential profiles appear to contain more dust,
show more recent star formation, are more flattened and more rotationally
supported (Kormendy 1992, Kormendy \& Kennicutt 2004, Fisher \& Drory 2008).
These objects are called pseudo-bulges, but one might also call them central
disks. For more detailed examples and references we refer the reader to the
comprehensive review by Kormendy \& Kennicutt (2004), which presents a complete
summary of observational evidence for this kind of bulge and their formation via
secular evolution: bulges could result from the evolution of disk dynamical
instabilities. Numerical simulations seem in fact to suggest that the
dissolution of bars inside the disks may trigger the formation of
three-dimensional stellar structures with roughly exponential profiles
(\citealt{pfenniger}, \citealt{combes}, \citealt{raha}, Norman, Sellwood, Hasan
1996). It is also known that these processes are more
effective in late-type galaxies rather than in early-types, despite the fact
that there are examples of pseudobulges in Sa and even S0 galaxies
(\citealt{beckman2003}, \citealt{kormendy04}).
In this paper, we will also adopt the photometric bulge-disk decomposition.

Our purpose in this paper is to investigate the nature and
interconnection of disks and bulges in a class of rather poorly studied objects:
late-type spiral galaxies, full of dust and star forming regions, and
characterised by relatively low-surface brightness; in the last years, galaxies
towards the end of the Hubble sequence have been the target for a number of
HST-based imaging surveys (\citealt{marcella97}, \citealt{marcella98a}, Carollo,
Stiavelli \& Mack 1998, \citealt{marcella99}, \citealt{marcella02},
\citealt{boker}, \citealt{laine}) which revealed the presence of a variety of
structures in their inner regions: bulges, nuclear star clusters, stellar disks,
small bars, double bars, star forming rings, whose formation and evolutionary
patterns are not properly understood yet. Because of the extinction, a
photometric analysis based on near-infrared data gives more reliable results
($A_H$ is roughly a factor 8 lower than $A_V$: Rieke \& Lebofsky 1985). Another advantage of
working in the NIR is that the NIR light is
much less affected by recent episodes of star formation than optical light, and
traces the old stellar populations hosting most of the luminous mass of
galaxies. For the analysis of
HST-NICMOS data the problem is, however, that the field of view is very small (about
20$''$ $\times$ 20$''$ for NIC2, so that an accurate determination of the sky
background is very difficult. Here we use data from the 2MASS survey to solve
this problem.  

The sample that we study here is special, in that accurate stellar kinematic
maps, stellar absorption line maps of a few strong lines, as well as emission
line strengths of [OIII] and H$\beta$ have been determined, making it possible
for us to study the relation between structural parameters and stellar
populations, the behaviour of ionized gas and the stellar potential. These maps
come from SAURON (Bacon et al. 2001) integral field spectroscopy. The stellar
and gas kinematics has been published in Ganda et al. (2006), while the
absorption line maps have been analyzed in Ganda et al. (2007). Our sample of
late-type galaxies can be easily compared to the early-type spirals of the
SAURON Survey (de Zeeuw et al. 2002), presented in Falc\'on-Barroso et al.
(2006) and Peletier et al. (2007). The fact that kinematic parameters are
available for this sample makes it possible to study the position of late-type
spirals on the M$_{Bulge}$ - $\sigma$ relations, establishing in this way
whether these bulges are similar in structure and formation to bulges of 
early-type spirals.

Given the renewed interest in the extinction in spiral galaxies (Driver et al.
2007, Graham \& Worley 2008, hereinafter GW) we decided to use the unique
property of this sample that the amount of extinction can be calculated by
combining absorption line maps, which are basically independent of extinction,
with broad band colour maps, such as the Mg~b maps, which are affected by it.
To do this, we determined high resolution $V-H$ maps for the 10 galaxies for
which HST-F606W ($V$)  images were available, in addition to F160W ($H$). The
combination of colour and line strength gives the colour excess E$_{V-H}$ in a
model-independent way, which can easily be converted to A$_V$. This method is
in principle very powerful, but has not been applied much in the literature,
because of the lack of well-calibrated line strength data.

The paper is structured as follows. Section \ref{samplesecch4} describes the
sample selection; Section \ref{profilesch4} the photometric profiles and the
methods applied for their extraction; Section \ref{bulgediscdecch4} the actual
bulge-disk decomposition; Section \ref{coloursecch4} compares the photometry in the
NIR with the optical $V-$ band, and presents extinction maps;
Section \ref{correlationsecch44} investigates the
correlations between the structural parameters and other galaxy properties;
Section \ref{centralcompsecch4} addresses the frequent presence of inner
additional components, and, finally, Section
\ref{conclusionsecch4} summarizes the main results.

\section[The sample]{THE SAMPLE}\label{samplesecch4}
%\mbremark{RC3 or LEDA? (They differ because the latter gets updated)}
The sample of galaxies on which we perform our bulge-disk decomposition is the
same for which we presented the two-dimensional kinematical maps from SAURON
integral-field spectroscopy in the paper by \citet{ganda} and the line-strengths
maps in the paper by \citet{ganda2}.\\ \indent The galaxies were optically
selected ($B_{T}$ $<$ 12.5, according to the values given in \citealt{RC3},
hereafter RC3) with HST imaging available from WFPC2 and/or NICMOS. Their
morphological type ranges between Sb and Sd, following the classification given
in NED (from the RC3). Galaxies in close interaction and Seyferts were
discarded. The resulting sample contains 18 nearby galaxies, whose properties
are listed and illustrated respectively in Table 1 and Fig. 1 in \citet{ganda}.

\section[Photometric profiles]{PHOTOMETRIC PROFILES}\label{profilesch4}

\subsection{Datasets used}
We constructed surface brightness profiles for all galaxies combining datasets
with different spatial resolution and field of view. We started from the 
$H-$ band images from the Two-Micron All Sky Survey (hereafter, 2MASS). 
%Since these data do not have a very high spatial resolution 
Since these data do not have a sufficient spatial resolution for our purposes
(2-3$''$; the pixel scale of the 
2MASS images
retrieved from the archive is 1\arcsec), we complemented the
photometric profiles extracted from the 2MASS images with profiles extracted
from NIR HST images (F160W), available for 11 out of 18 cases (marked with a
single or double dagger in Table \ref{tab1ch4}). To improve the spatial
resolution at small radii for the remaining galaxies we used optical HST
images (F814W, corresponding to the $I-$ band). Since, on the other hand,
2MASS is rather shallow (reliable profiles can be extracted
out to $\approx$ 5 - 8 kpc for most of our galaxies), we needed an additional
dataset for the outer parts, so that the disk geometry could be determined
unambiguously, and an accurate sky level could be determined. 
For this
reason, we decided to complement the profiles also with profiles extracted from
infrared\footnote{We used POSS-II Near-IR IVN+RG9 images, obtained from
http://archive.stsci.edu/cgi-bin/dss\_form/; the selected filter is close to the
$Z-$ band.} images from the Digitized Sky Survey (hereafter, DSS), which are
very deep, and therefore radially very extended. This also allows us a more
accurate determination of the sky background. The DSS, 2MASS and NICMOS images
downloaded from the archives were fully reduced, so we did not apply any further
processing; as for the WFPC2-F814W images, we first created a mosaic from the
four chips in order to maximize the field of view, using the {\small IRAF} task
{\small WMOSAIC} in the {\small STSDAS.HST\_CALIB.WFPC} package, rotated the
resulting image to orient it to North up-East left and performed cosmic rays
removal, using the {\small IRAF} tasks {\small CRREJ} and {\small LACOS\_IM},
created by Pieter van Dokkum and available via his
website\footnote{http://www.astro.yale.edu/dokkum/lacosmic/}.

\subsection{Ellipse fitting}
As a first step, we extracted the photometric profiles from the DSS images using
the {\small IRAF} task {\small ELLIPSE} in the {\small STSDAS.ANALYSIS.ISOPHOTE}
package, which fits elliptical isophotes to galaxy images, implementing the
method initially described by Kent (1983, 1984) and \citet{jed}. 
We masked out bad pixels and foreground stars, but did
not exclude dust lanes and star forming regions from the fitting. We first
fitted ellipses to the images with the centre, position angle and ellipticity
left as free parameters. In this way we obtained profiles of the centre
coordinates; from these profiles we established the position of the centres,
kept fixed in the following steps. In case of ambiguity, the centre was
determined using isophotes at intermediate radii, since the very central ones
can be affected by dust obscuration, and, in the particular case of the DSS
images, by the seeing (or even by saturation of the images, in the worst cases).
\\ \indent We then fitted again ellipses
to the images with the centre fixed at the chosen position and ellipticity and
position angle free. From this second step we determined single values for the
geometric parameters. In many cases in the outer parts of the galaxies, outside
the nuclear region, the position angle and ellipticity converge to more or the
less constant values. In other cases the situation is more complicated,
especially for galaxies that are almost round, and there we picked the values
that looked more reasonable at a visual inspection of the shape of the isophotes
in the images.

\begin{table}
\begin{center}
 \begin{tabular}{@{}l | l l l l l l l l}
\hline \hline
NGC & Type & M$_B$ & d & Scale & PA& $\epsilon$& PA$_{lit}$ & $\epsilon_{lit}$\\
(1) & (2) & (3) & (4) & (5) & (6) & (7) & (8) & (9) \\
\hline
488$\dagger$  &3.0 &-21.71  &32.1 &  156 & 5  &      0.23  &   15   &	0.260\\
628$\dagger$  &5.0 &-20.29  &9.8  &  47  & 25  &  0.19   &  25   &   0.088\\
772$\dagger$  &3.0 &-22.23  &35.6 &  173 & 126  & 0.34  &   130  &   0.411\\
864$\dagger$  &5.0 &-20.54  &21.8 &  106 & 26  &  0.32   &  20  &    0.241\\
1042          &6.0 &-19.83  &18.1 &  88  & 174 &  0.29   &  43   &   0.224\\
2805          &7.0 &-20.75  &28.2 &  137 & 125 &   0.24  &   125  &   0.241\\
2964$\dagger$ &4.0 &-19.74  &20.7 &  100 & 96  &  0.45   &  97  &    0.451\\
3346          &6.0 &-18.89  &18.9 &  92  & 100 &  0.16  &   111  &   0.129\\
3423          &6.0 &-19.54  &14.7 &  71  & 41 &      0.23  &   10  &	0.149\\
3949$\dagger$ &4.0 &-19.60  &14.6 &  71  & 122 &  0.36  &   120  &   0.425\\
4030$\dagger$ &4.0 &-20.27  &21.1 &  102 & 37  &  0.24  &   27   &   0.276\\
4102$\dagger$ &3.0 &-19.38  &15.5 &  75  & 42  &  0.445  &  38   &   0.425 \\
4254$\dagger$ &5.0 &-22.63  &19.4 &   94 & 50  &  0.27   &  62   &   0.129\\
4487          &6.0 &-19.12  &14.7 &  71  & 77  &  0.37  &   75       &  0.324\\
4775          &7.0 &-19.81  &22.5 &  109 & 96  &  0.135 &   52  &    0.067\\
5585$\ddagger$&7.0 &-18.32  &8.2  &  40  & 38  &  0.36   &  30  &    0.354\\
5668          &7.0 &-19.65  &23.9 &  116 & 120 &  0.155  &  164  &   0.088\\
5678$\dagger$ &3.0 &-21.30  &31.3 &  152 & 5   &  0.475 &   5	&    0.51\\
\hline
 \end{tabular}
\end{center}
\caption[Ellipticities and position angles for the 18 galaxies.]{Column (1): NGC
identifier; column (2): morphological type (RC3); column (3): absolute blue magnitude,
from HyperLeda, computed using distance in Mpc given in column (4); column (5): scale 
in pc/arcsec;
columns (6) and (7): adopted position angle (PA) and
ellipticity ($\epsilon$) for the measurement of the photometric profiles (N-->E);
columns (8) and (9): position angle and ellipticity values tabulated in the
literature (PA$_{lit}$ and $\epsilon_{lit}$); we refer to the RC3 values, with
the exceptions of NGC\,1042, 3346, 4254, 4775, 5668, where the angle is taken
from \citet{grosbol}. The galaxies marked with a single or double dagger are
those for which we used NICMOS imaging from the NIC2 or NIC3 cameras,
respectively. For the remaining galaxies WFPC2-F814W images are available.}
\label{tab1ch4}
\end{table}

Table \ref{tab1ch4} lists the position angles and ellipticities of our sample
galaxies, together with some other basic data, and shows that in most cases the
chosen values are very close to the values tabulated in the literature (mainly
in the RC3);  the cases which present
the biggest discrepancies are actually galaxies almost face-on and/or very
round, for which the errors on the geometric parameters are in any case big.
Once we had determined the geometric parameters, using {\small ELLIPSE} again we
measured the intensities at each radius with centre coordinates, position angle
and ellipticities fixed for all ellipses. From the resulting photometric
profiles we subtracted then a value for the sky background estimated on the
outer parts of the images. Following this, 
we fitted ellipses to the 2MASS images, with all the geometric parameters
(centre coordinates, position angle and ellipticity) free and determined the
centre coordinates in the same way as done for the DSS images. And after this we
measured the intensity profiles, fixing the centre coordinates to the values 
just
determined and the position angle and ellipticity to the values determined on
the basis of the DSS profiles and subtracted an estimated value for the sky
brightness. An analogous procedure was repeated for the NICMOS images, when
available, and for the WFPC2-F814W images, in the cases where NICMOS was missing
(see Table \ref{tab1ch4}).

\subsection{Combining the profiles}
At the end of this procedure, for each galaxy we combined the three photometric profiles,
to get a single profile with the maximum radial extension and spatial resolution 
allowed by the data. To do this, we first
combined the profiles extracted from the DSS and 2MASS images, aligning them at
the 2MASS level, and transformed the resulting `ground-based' profile to an
absolute magnitude scale using the zero point calibration (the keyword MAGZP) of
the 2MASS images. For the 11 galaxies with available
NICMOS imaging, we first calibrated independently the NICMOS profile to the
absolute $H$ (Vega) magnitude scale using the F160W zero points of 21.826 (plus 
a term taking into account the pixel scale)for the 10
galaxies with NIC2 data and
21.566 for NGC 5585, observed with NIC3, and then joined
them with the calibrated `ground-based' profiles; the quality of the matching of
the two profiles gave us an indication of the quality of our estimation of the
sky background in the HST image, which was then changed iteratively until a
satisfactory result was obtained. For the remaining galaxies, we imposed as a
photometric zero point to the HST profiles (F814W) the mean offset between the
aperture magnitudes measured on the 2MASS and HST images in the radial range
7-12\arcsec, which depends on an estimate of the sky background level for the
HST image; we then joined the HST and the calibrated `ground-based' profiles,
modifying our sky estimate for the HST image until a good match was obtained;
given the fact that the field of view of WFPC2 is significantly larger than the
one of NICMOS, this operation required much less careful fine-tuning than for
the NICMOS profiles. At the end, calibrated global $H-$ band profiles were obtained. 
We tested whether the use of WFPC2-F814W images in the cases where NICMOS is not
available introduces significant additional errors: for several of the
galaxies with available NICMOS imaging, we retrieved the F814W images and
treated them in the way described above; the result is that the comparison
between the final global profiles obtained this way and the ones obtained using
NICMOS is acceptable, with differences below 0.1 mag outside the inner $\approx$
1\farcs5 and below 0.05 mag outside the inner $\approx$ 4\farcs5. This means
that we can be rather confident in using the F814W images for the inner regions,
and that $I - H$ colour gradients are not very important. 

\section[Bulge-disk decomposition]{BULGE-disk DECOMPOSITION}
\label{bulgediscdecch4}
\label{lumsec}
Fitting a parametric bulge and disk can be done directly on the 2-dimensional
images, or on one-dimensional profiles, that have been obtained by averaging
azimuthally in ellipses or circles. Although one-dimensional profiles suffer
from loss of information due to projection, and produce larger degeneracies
between bulge and disk parameters (\citet{jong96bis}), two-dimensional fits are
less robust, and degeneracies remain (\citealt{mollenhoff}). 
On the basis of their simulations \citet{lauren} conclude that a full
two-dimensional technique does not provide a significant improvement compared to
one-dimensional methods in recovering the axysimmetric structural parameters.
Based on this, we have chosen an
approach that allows us not to have to assume the same shape of bulge and disk, 
but for which we do not have to apply a full two-dimensional fitting method, 
as we will explain in the
following sections. Alternatively, we could have used a non-parametric fitting
method, such as the method of Kent (1984), where two components are found with
different axis ratios, but have not done this, because of the degeneracies that
arise when choosing the axis ratios of these components.

\subsection{Decomposing late-type spirals}
We performed a non-simultaneous bulge-disk decomposition for our 18 galaxies,
closely following the method described by \citet{edo} and \citet{edo2007} and
previously introduced by \citet{palu}. In a few words, we fit an exponential law
to the disk, marking the fitting range via a visual inspection of the profiles,
choosing the part of the profile that is exponential without doubt; we build a 
model image of the disk and subtract it from the original images, obtaining
`bulge images' from which we extract the `bulge profiles' that we then fit with
a S\'ersic $r^{1/n}$ law.

\subsubsection{disk fitting}\label{discfittingfirstch4}
\begin{figure*}\centering
{\includegraphics[width=\textwidth]{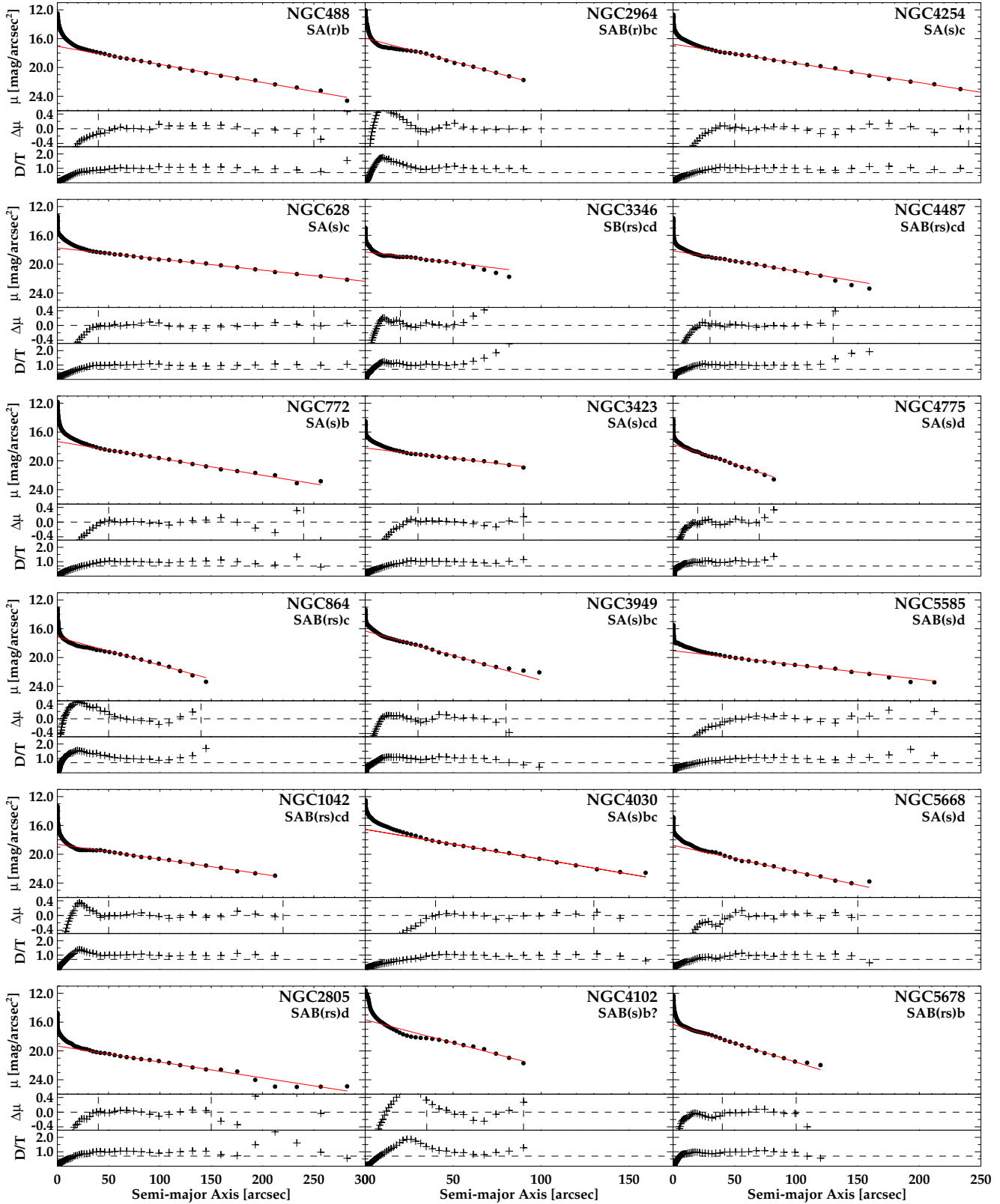}}
\caption{Global combined photometric profiles and exponential fits to the disks.
Filled dots show the global combined photometric
profiles, along with the best-fitting exponential (solid line).
The units are arcsec along the semi-major axis of the ellipses on the horizontal
axis and absolute $H$ magnitudes arcsec$^{-2}$ along the vertical. Below each
profile the residuals from the fit are shown, in the same units. 
Below this, the disk to total ratio of the intensity of the fitted exponential 
disk and the total profile intensity is shown; 
the level $D/T = 0.7$ is indicated
(dashed line), defining the `bulge region', as explained in the text.}
\label{fig6ch41}
\end{figure*}

The first step in our bulge-disk decomposition is the estimation of the disk
parameters by means of profile fitting. On the combined global profiles,
measured as explained in Section \ref{profilesch4}, using a least-square
algorithm we fitted an exponential light distribution (in mag~arcsec$^{-2}$):

\begin{equation}
\mu_{d}(r)=\mu_{0,d}+1.0857\left(\frac{r}{h}\right)
\end{equation}

where $\mu_{0,d}$ and $h$ are the central 
surface brightness and scale length of the disk. The actual fit was performed on
the profiles in absolute magnitudes and on radial ranges chosen by means of a
visual inspection, in order to avoid the contamination from the central
component (the `bulge'). The choice of a fitting range became a sort of
`compromise' in a few cases where the shape of the profile would have suggested
the presence of a double disk (see \citealt{pohlen}). 
Fig. \ref{fig6ch41} shows the
global galaxy profiles, together with the best-fitting exponentials, 
and the residuals from the fit. The radial
ranges of the fits are indicated in the residual plot. 
These figures show that in general the fits
are satisfying. NGC\,4102 represents the case where the fit is worst; this might
be due to contamination from the big bar, or from the coexistence of an inner
and an outer disk (see also \citealt{pohlen}), to which we impose a
single-exponential fit.

The fit parameters are listed in Table \ref{tab2ch4}. No
correction for inclination or galactic extinction is applied.
The uncertainties of these parameters can be calculated in various ways. Often
uncertainties in the least-squares fit are quoted. These are often very small,
since galactic disks cane be fitted very well with an exponential surface 
brightness profile, if a limited radial fitting range is used (de Vaucouleurs
1959, Freeman 1970, MacArthur et al. 2003, Pohlen \& Trujillo 2006). By 
changing the radial range, however, disk parameters can vary considerably.
Therefore, a more realistic estimate of the uncertainties in surface brightness
and scale length is obtained if one studies the variation of these parameters
as a result of changing the fitting range. To do this, we varied both the inner
and outer radial limit by steps of 5\%, from 70\% to 130\% of the range quoted
in Table 2. The final parameters disk are the average parameters obtained in
this way, and the uncertainty the RMS scatter.

\subsubsection{Bulge fitting}\label{bulgefittingch4}
\renewcommand{\thefigure}{\arabic{figure}}

In all cases, in the innermost regions the galaxy's light clearly exceeds the
exponential fit to the disk. According to the photometric definition (see the
Introduction), the region of the galaxy corresponding to this light excess is
the bulge. We will adopt this definition, but without implications on the
kinematics and/or populations: without implying that it has to be a hot and old
component, similar to a {\it small elliptical}.

To quantify the region where the light from the bulge is significant, we
established a {\it bulge radial extension}, defining as {\it bulge} the region 
within
which the intensity ratio between the fitted exponential disk and the total
intensity is lower than 0.7 (i.e. the bulge contributes to  30 percent or more
of the total light, see Fig. \ref{fig6ch41} and Table \ref{tab2ch4}). Note that
these figures can be affected by the presence of bars or other
asymmetric features not belonging to the exponential disk. 

\begin{figure}\centering
{\includegraphics[width=0.99\linewidth]{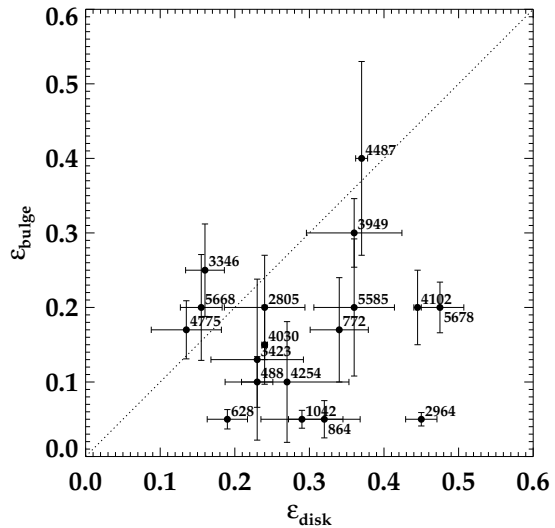}}
\caption{Bulge against disk ellipticities; each filled dot represents a galaxy,
with the NGC number indicated close to the symbol. The disk ellipticity is
determined from the outer parts of the galaxies, the bulge ellipticity from the
inner regions, as explained in the text. The dotted line represents the 1:1
relation. The overplotted error bars are
obtained from the scatter of the points in the profiles.}
\label{fig7ch4}
\end{figure}

On the basis of the fitted disk parameters, using the standard {\small IRAF} tasks
({\small BMODEL} in the {\small STSDAS.ANALYSIS.ISOPHOTE} package) we built a
model image of the disk and subtracted it from the original 2MASS images. We
will refer to the residual images as the 2MASS `bulge image'.
From these images we extracted the bulge photometric profiles and parameters, by
applying a procedure similar to the one used to measure the `total' profiles as
explained in Section \ref{profilesch4}: we fitted ellipses to the bulge images
in two steps, keeping in all cases the centre fixed to the value already
determined for the original 2MASS images and the position angle fixed to the
value determined from the DSS images in Section \ref{profilesch4}. At first, we
left the ellipticity free; in a second step we fixed also the ellipticity and
extracted the intensity profiles from the 2MASS bulge images, successively
converting them to an absolute magnitude scale, applying the offset given by the
keyword MAGZP in the 2MASS headers. As ellipticity for the bulge we adopted a
representative value estimated within the bulge region, on the basis of the
ellipticity profiles extracted from the 2MASS and HST images. 
We notice that bulge and disk ellipticities can be different. Fig. \ref{fig7ch4}
presents a comparison between the adopted values for the ellipticities of disks
and bulges, showing that for most of the galaxies the inner component is rounder
than the disk; but in some cases (NGC\,3346, 4487, 4775, 5668, among the
latest-type objects in our sample), the bulge is instead flatter than the disk,
as shown also by \citet{kambiz} on the basis of the bulge-disk decomposition of
70 disk galaxies spanning a range in type between S0 and Sm. Only NGC\,3346
lies more than 1$\sigma$ 
%away from 
above the 1:1 relation in
Fig. \ref{fig7ch4}. 

Once the bulge and disk profiles have been retrieved, we need to determine the 
structural parameters by fitting a model profile. We follow the approach most
often used in the current literature (see Graham \& Worley 2008 for a
comprehensive review) by fitting an exponential disk and a S\'ersic $r^{1/n}$
bulge.

\begin{figure*}\centering
{\includegraphics[width=\textwidth]{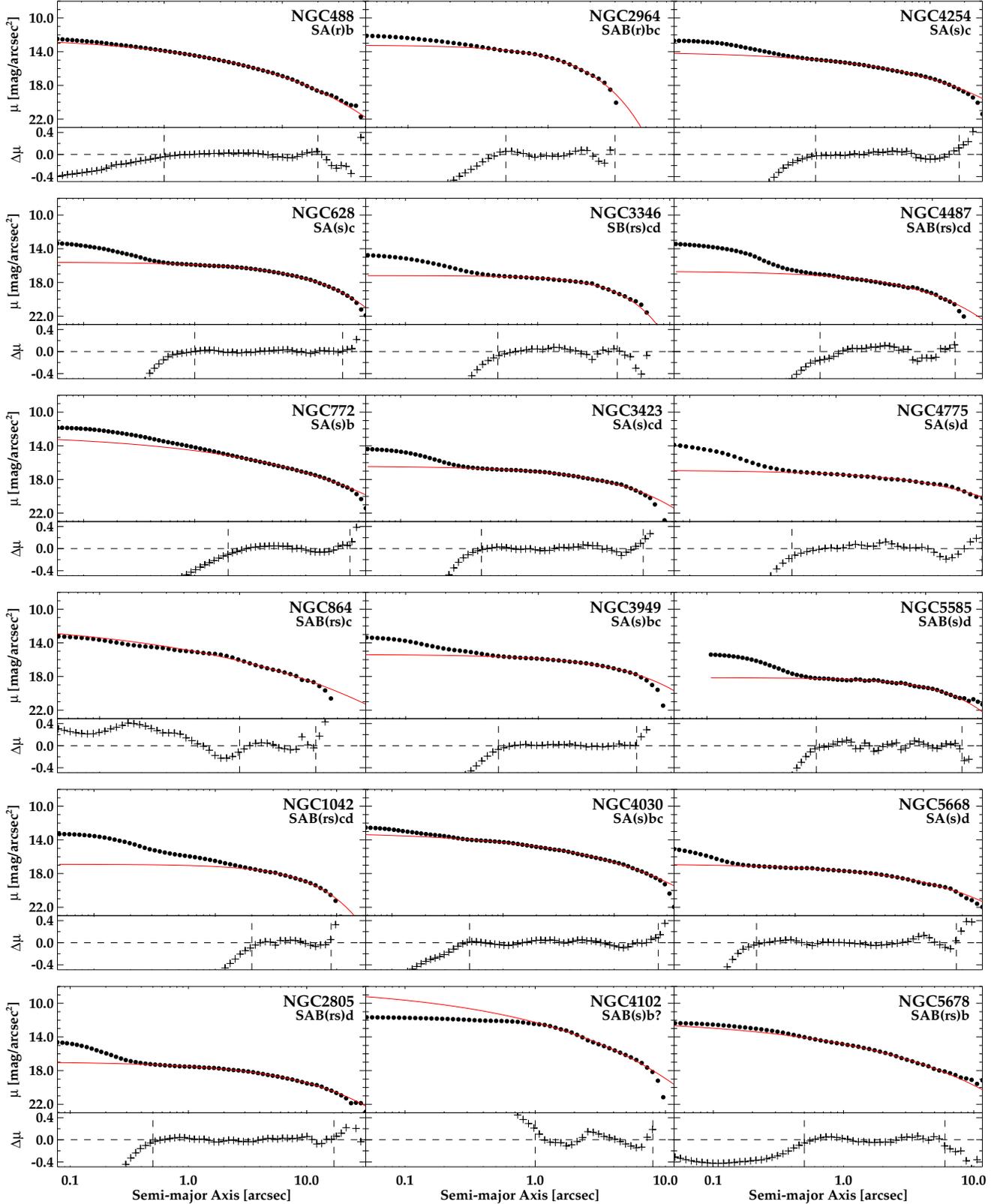}}
\caption{S\'ersic fits to the bulge profiles. In the top part of each panel we
plot the HST or combined HST + 2MASS bulge profiles (filled dots), in mag
arcsec$^{-2}$; the NICMOS images are used in all available cases, the
F814W-WFPC2 in the others (see Table \ref{tab1ch4}); overplotted are the
best-fitting S\'ersic profiles (solid line);  The low part of each panel shows 
the residuals from the fit (in mag
arcsec$^{-2}$), as well as the fitting range (vertical lines). Note that the 
abscis is given in logarithmic units, contrary to Fig. \ref{fig6ch41}. }
\label{fig8ch4}
\end{figure*}

When fitting the profiles,
seeing effects are particularly relevant when the effective radius is small
and/or when the ratio between the effective radius and the FWHM of the seeing is
small \citep{graham}. In fact, for small bulges poor seeing could smear the
images so that bulges intrinsically described by $n >$ 1 S\'ersic profiles could
appear like $n <$ 1 profiles. Therefore, for a reliable treatment of the seeing
effects, we performed the fit on the {\it bulge region} by convolving the 
S\'ersic
profiles with a Gaussian point-spread function, as explained by \citet{graham},
and then fitting the seeing-convolved S\'ersic profiles using a standard
non-linear least-squares algorithm. The values of the FWHM of seeing in the
2MASS images were retrieved from the header keyword SEESH via the relation
FWHM(\arcsec) = 3.13 $\times$ SEESH $-$ 0.46 as explained on the 2MASS
webpage\footnote{www.sao.arizona.edu/FLWO/2mass/seesum.html}. In some cases
(e.g. NGC\,864, 2964, 3346, 3949, 4775 and 5678) the bulge region is so small that
it is comparable in size with the FWHM of seeing in the 2MASS images (less than 3
$\times$ FWHM of the seeing), so that the applied procedure is not
meaningful and possibly leads to imprecise fitting  values. In order to face
this problem, to gain spatial resolution in the inner parts, and to take advantage of
data non contaminated by seeing effects, we perform the bulge fitting
on bulge profiles extracted from the HST images (NICMOS or F814W, depending on
availability). These profiles were obtained by subtracting from the HST images
an exponential disk, modeled on the basis of the fit parameters obtained as
described in Section \ref{discfittingfirstch4}, and then fitting to the residual
images ellipses with position angle and ellipticity fixed to the `disk position
angle' and the `bulge ellipticity', and centre coordinates fixed as well. Some
of these `HST bulge profiles' suffer from the complementary problem of limited
spatial extension: in several cases they do not cover the full bulge region.  In
those cases (NGC\,488, 628, 772, 4030, 4254, 4775, 5668) we joined them with the
bulge profiles previously obtained from the 2MASS images. 
On these profiles we fitted a
pure S\'ersic law, excluding the very innermost parts (0.5-1\arcsec), which in
several cases host nuclear star clusters. In Fig. \ref{fig8ch4} we show the HST
or joined HST + 2MASS bulge profiles, along with the best-fitting S\'ersic profile,
and
the residuals from the fit. 
%We find $n$ values ranging between $\approx$ 0.7 and $\approx$ 3.

\begin{figure*}\centering
{\includegraphics[width=0.99\textwidth]{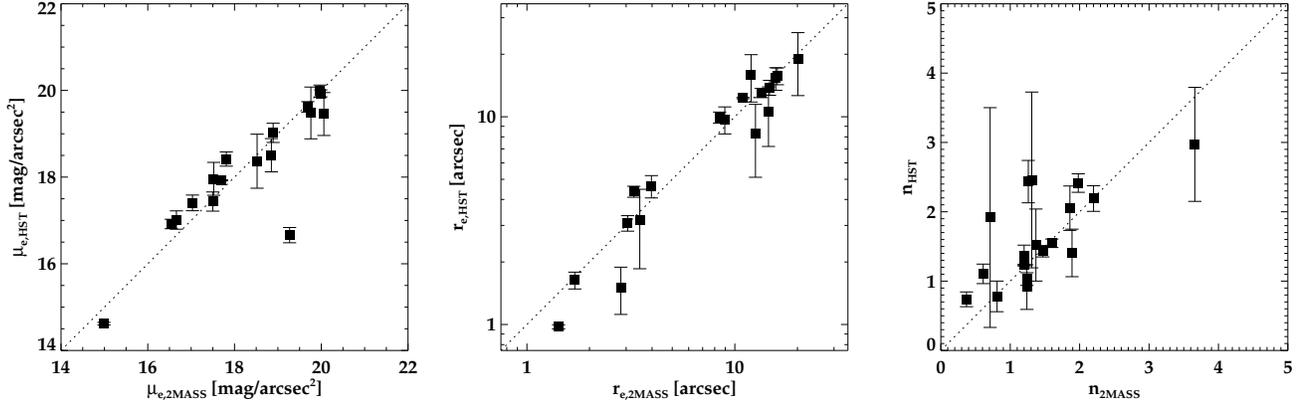}}
\vspace{0.3cm}
\caption{Comparison between the fitted parameters for a seeing-convolved
S\'ersic fit performed on the 2MASS bulge profiles and a
pure-S\'ersic fit on the HST or joined HST + 2MASS bulge profiles, excluding the
innermost region: from left to right, effective surface brightness, effective
radius, $n$. The dotted line in each panel represents the 1:1 relation.}
\label{fig9ch4}
\end{figure*}

\begin{table*}
\centering
\small
\tabcolsep=1.3mm
 \begin{tabular}{@{}l | c c c c c c | c c c c c c c c c c c  | c c c}
\hline \hline
NGC & r$_{d,i}$ & r$_{d,o}$ & $\mu_{0,d}$ & $\pm$ & h & $\pm$ &
r$_{bu,i}$ & r$_{bu,o}$ & $\mu_{e,2m}$ & $r_{e,2m}$ & $n_{2m}$ & $\mu_{e,hst}$ & $\pm$ & $r_{e,hst}$ & $\pm$ &
$n_{hst}$ & $\pm$ & M$_{d}$ & M$_{bul}$ & $\pm$ \\
(1) & (2) & (3) & (4) & (5) & (6) & (7)& (8) & (9) & (10)& (11) &(12) & 
(13) & (14) & (15) & (16) & (17) & (18) & (19) & (20) & (21)  \\
\hline
488  & 40 &  250 & 17.03 & 0.07 & 42.7 & 1.1 &1  & 20 & 16.54	& 8.5 & 1.98  & 16.92 & 0.11& 9.9& 0.6 & 2.41 & 0.13 &-25.70 &   -23.77 & 0.05 \\
628  & 40 &  250 & 17.75 & 0.03 & 70.7 & 1.1 &1  & 21.5 & 17.70 & 10.9 & 1.20  &17.92 & 0.01&12.3& 0.1 & 1.23 & 0.01 &-23.50 &   -20.31 & 0.01 \\
772  & 50 &  240 & 17.35 & 0.06 & 47.1 & 1.2 &2  & 25 & 18.52	& 20.1 & 3.66  &18.37 & 0.62&19.0& 6.4 & 2.9 & 0.8   &-25.82 &   -24.07 & 0.25 \\ 
864  & 50 &  140 & 17.13 & 0.30 & 28.1 & 2.7 &1  & 4  & 16.66  & 1.7 & 0.71    & 17.01 & 0.21& 1.6 & 0.2 & 1.9 & 1.6 &-23.85 &   -18.78 & 0.42 \\
1042 & 50 &  220 & 18.62 & 0.02 & 52.5 & 0.7 &2  & 9 & 17.81	& 3.3 & 1.24   & 18.42 & 0.16& 4.4& 0.3 & 0.91 & 0.32&-23.32 &   -18.75 & 0.14 \\
2805 & 40 &  150 & 19.36 & 0.06 & 50.8 & 1.8 &0.5& 17 & 19.99	& 13.4 & 1.60  & 19.93 & 0.08&13.0& 0.7 & 1.55 & 0.06&-23.46 &   -20.84 & 0.05 \\ 
2964 & 30 &  100 & 15.96 & 0.07 & 16.9 & 0.3 &0.5& 3 & 14.99	& 1.4 & 0.37   & 14.62 & 0.03& 1.0& 0.0 & 0.74 & 0.11&-23.79 &   -19.47 & 0.04 \\
3346 & 20 &  50  & 18.28 & 0.14 & 35.4 & 4.5 &0.5& 4 & 18.85	& 3.5 & 0.81   & 18.51 & 0.38& 3.2& 1.3 & 0.78 & 0.22&-22.90 &   -17.99 & 0.35 \\
3423 & 30 &  90  & 18.28 & 0.08 & 40.1 & 2.5 &0.5& 12.5 & 18.89 & 8.9 & 1.20   & 19.02 & 0.22& 9.7& 1.4 & 1.37 & 0.15&-22.63 &   -19.64 & 0.15 \\
3949 & 30 &  80  & 16.48 & 0.29 & 16.8 & 1.6 &0.5& 5.5 & 17.03  & 4.0 & 0.61   & 17.41 & 0.18& 4.6& 0.6 & 1.10 & 0.14&-22.51 &   -19.52 & 0.13 \\
4030 & 40 &  130 & 16.55 & 0.09 & 26.5 & 0.8 &0.5& 25 & 17.50	& 15.7 & 2.20  &17.44 & 0.22&15.3& 1.9 & 2.19 & 0.18 &-24.22 &   -23.23 & 0.10 \\
4102 & 35 &  90  & 16.03 & 0.45 & 19.5 & 3.1 &1  & 8.5 & 14.12  & 2.8 & 1.31   & 12.99 & 0.62& 1.5& 0.4 & 2.5 & 1.3  &-23.40 &   -22.00 & 0.25 \\
4254 & 50 &  240 & 16.76 & 0.04 & 40.7 & 0.6 &1  & 18 & 17.52	& 12.0 & 1.86  &17.96 & 0.38&15.8& 4.1 & 2.05 & 0.32 &-24.78 &   -22.57 & 0.20 \\
4487 & 30 &  130 & 18.07 & 0.12 & 36.8 & 2.4 &1  & 10 & 19.76	& 14.5 & 1.37  &19.48 & 0.60&10.6& 3.4 & 1.52 & 0.52 &-22.64 &   -19.41 & 0.30 \\
4775 & 20 &  70  & 17.67 & 0.06 & 19.3 & 0.6 &0.5& 6.5 & 20.06  & 12.6 & 1.89  &19.46 & 0.49& 8.3& 3.2 & 1.41 & 0.34 &-22.55 &   -19.79 & 0.37 \\
5585 & 40 &  150 & 19.00 & 0.09 & 53.2 & 2.9 &1  & 21.5 & 19.97 & 16.0 & 1.24  &19.98 & 0.14&15.7& 1.5 & 1.03 & 0.09 &-21.26 &   -18.33 & 0.10 \\
5668 & 40 &  150 & 18.80 & 0.16 & 30.3 & 1.7 &0.5& 18 & 19.69	& 14.7 & 1.47  &19.62 & 0.12&13.8& 1.1 & 1.43 & 0.08 &-22.54 &   -20.89 & 0.09 \\
5678 & 40 &  100 & 16.34 & 0.25 & 20.9 & 1.7 &0.5& 6 & 19.27	& 3.1 & 1.26   & 16.66 & 0.18& 3.1& 0.3 & 2.43 & 0.31&-24.78 &   -21.42 & 0.08 \\
\hline
 \end{tabular}
\caption[\small Fit parameters and structural parameters from our photometric
bulge-disk decomposition.]{\small Column (1): NGC identifier; columns (2) and
(3): disk fitting range, in arcsec; columns (4) and (6): respectively, central surface
brightness of the disk in mag arcsec$^{-2}$ and scale length of the disk in
arcsec, from the fits performed on the photometric profiles in Section
\ref{discfittingfirstch4}; column (5) and (7) indicate their uncertainties;
columns (8) and (9): bulge fitting range, in arcsec;
 columns (10), (11) and (12): 
effective surface brightness in mag arcsec$^{-2}$, effective radius in arcsec
and S\'ersic parameter ($n$) of the bulge from the fit of a seeing-convolved
S\'ersic profile to the bulge profiles extracted from the 2MASS images; columns
(13), (15) and (17): bulge parameters (effective surface brightness, effective
radius and S\'ersic parameter ($n$) from the fit of a S\'ersic profile to the
HST or joined HST + 2MASS bulge profiles, with uncertainties in columns (14),
(16) and (18); columns (19) and (20): $H-$ band absolute magnitudes of the disk
and bulge components, and the uncertainty in the latter (21). The
uncertainties quoted here are the variations due to changing the fitting
ranges.  }
\label{tab2ch4}
\end{table*} 

The fit parameters for all galaxies, for both the disk and the bulge components,
are listed in Table \ref{tab2ch4}. Here we report the disk parameters from the
exponential fit as in Section \ref{discfittingfirstch4}; for the bulge we quote
the parameters from the seeing-convolved S\'ersic fit to the 2MASS bulge
profiles and those from the S\'ersic fit to the HST or combined 2MASS-HST bulge
profiles.  No correction for inclination or galactic extinction is applied to
the surface brightness (and magnitudes) values reported in the table. 
The uncertainties of the bulge parameters have been calculated in the same 
way as for the disk, i.e. by varying the fitting range by 30\% for both the 
inner and outer boundary. Fig.
\ref{fig9ch4} shows the comparison between the two sets of parameters for the
bulge; the effective surface brightness and radius are generally rather similar,
while the S\'ersic parameter $n$ tends to be larger for the fits on the HST (or
combined HST + 2MASS) profiles. In the following, for the bulge parameters we
will refer to the output of the fit on these latter profiles.
From the parameters we computed the total bulge and disk luminosities (see Graham
2001), and 
converted them to absolute magnitudes M$_{d}$ and M$_{b}$ using the distance moduli
used by \citet{ganda}. These numbers, together with the bulge to disk ratio, 
are also given in 
Table \ref{tab2ch4}. 

\section[Optical photometry and optical $-$ NIR colours]{OPTICAL PHOTOMETRY AND OPTICAL $-$ NIR
COLOURS}\label{coloursecch4}
As we mentioned in the Introduction and as we witnessed throughout this work,
late-type galaxies are known to be dusty objects, with dust lanes and structures
extending often all the way to the centre (\citealt{frogel}, Zaritsky, Rix \& Rieke
1993, Martini et al. 2003). For example we just mentioned in the previous
section that dust makes it sometimes difficult  to investigate the properties of
tiny inner components. This is one of the reasons for which we decided to work
in the near-infrared.  In order to investigate the amount and distribution of
extinction in the centers of these objects,  we decided to analyse also optical
images. Here, the fact that we have absorption line strength maps determined
with SAURON offers a large advantage, when compared to previous work. Comparing
optical-near infrared colours, affected by extinction, with absorption line
index maps, which are almost not affected, we can obtain a reasonably accurate
guess of the amount of extinction in the central regions of a representative
sample of late type spirals. There are 13 galaxies for which F606W HST images
are available in the HST archive. For 10 of them, also NICMOS F160W images are
available. We decided to analyse these 10 galaxies.  

\subsection{Optical -- Near-infrared colour profiles}
We analyzed the optical images in the same way as the NIR ones, with the exception that here we
used the POSS-II Red IIIaF + RG610 images in the outer parts: using the {\small IRAF} task
{\small ELLIPSE} we first extracted  from the DSS images photometric profiles with centres and
geometric parameters free and determined the centres coordinates; we then fixed the centres and 
extracted again profiles with free ellipticities and position angles. In many cases in the outer
parts the geometric parameters converge to a  constant value; in the majority of cases this
value is comparable with what found in the NIR: we did not detect any significant difference in
the  shape of isophotes between optical and near-infrared. Therefore, we fixed the position
angle and ellipticity to the values determined on the NIR images  (see Table \ref{tab1ch4}) and
measured the intensities along the ellipses. We then subtracted the sky background, estimated by
averaging the mean values of intensity  in several boxes placed in the outer parts of the
images, and converted the profiles to a (uncalibrated)  magnitude scale. For the HST F606W
images, we started  by creating mosaic images of the four chips, using the {\small IRAF} task
{\small WMOSAIC}, rotated them to bring them to the standard orientation with North up-East
left  and extracted from them intensity profiles with centre, position angle and ellipticity
fixed (fixed to the NIR values); we then subtracted an  estimate for the sky background and
converted the profiles to a (still uncalibrated) magnitude scale. Then we combined HST and DSS
profiles into single profiles, aligning them at the level of the HST profiles; when a good
matching of the two was not achieved, we modified the estimated sky brightness in the HST
image,  until the matching was satisfactory. The last step is the photometric calibration:
converting the profiles to an absolute magnitude scale. We calibrated our profiles  to the $V-$
band of the UBVRI Johnson-Cousin system using the equations provided by \citet{holtzman}, that
were derived for single-chip images. To be able  to apply those transformations, for each galaxy
we focused on the single-chip image containing the galaxy centre (PC1 for all galaxies excluding
NGC\,1042 and 3949,  centred instead on WF3) and extracted the single-chip profiles. The sky
subtraction is in this case more difficult, since the field of view is significantly smaller:
we  proceeded in an iterative way, changing the estimated value of the sky background until the
sky-subtracted single-chip profile matched the shape of the  sky-subtracted profile retrieved
from the mosaic image. We then calibrated the single-chip profiles applying Equation 9 in
\citet{holtzman}, using the coefficients listed  in his Table 10:
%\begin{center}
\begin{displaymath}
\mu_{V} = - 2.5\, \times \log(counts) \,+ 2.5\, \times \log(EXPTIME) \,\,\,+ 
\end{displaymath}
\begin{equation}
\,\,\,\,\,\,\,\,\,\,\,\,\,\,\,\, + \,5 \times \log(SCALE) \,+ \,0.254 \times (V - I) + \,0.012\, \times  
\end{equation}
\begin{displaymath}
\,\,\,\,\,\,\,\,\,\,\,\,\,\,\,\, \times (V - I)^{2} \,+22.093 \,+\, 2.5 \times \log(GR) \,+\, 0.1
\end{displaymath}
%\end{center}
where `$counts$' is the number of counts in linear units, `$EXPTIME$' the integration time on the image, `$SCALE$' the pixel scale (arcsec pixel$^{-1}$, $\approx$ 0.046 for the 
galaxies centred on PC1 and $\approx$ 0.09944 for the two centred on WF3), $V - I$ the colour in the
Johnson-Cousins UBVRI system, 22.093 the photometric zero point tabulated by \citet{holtzman}, $GR$ equals 2.003 for frames where a gain of 14 was used, and 1 otherwise, and 0.1 is added, following the prescription of \citet{holtzman}, in order 
to correct for infinite aperture (see also the WFPC2 Data Handbook). For the ($V - I$) colours of our galaxies we used values from
HyperLeda and, when not
available, an average of the values for galaxies of the same Hubble type. Since the colour term is relatively small, there is no need for extremely accurate values here. The adopted values are listed in Table \ref{tab6ch4}. The last step was to find and apply the vertical 
offset necessary to align the HST(mosaic) + DSS joined profiles to the calibrated single-chip profiles.\\
\begin{table}\centering
\small{\begin{tabular}{@{}l l | l l }
\hline \hline
NGC &  $ V - I$ &NGC &  $ V - I$  \\
\hline
488 & 1.16  &3949 & 0.81 \\ 
628 & 1.38  &4030 & 1.20 \\ 
864 & 1.28  &4102 & 1.23 \\
2964 & 1.28 &5585 & 0.88 \\ 
3949 & 0.81 &5678 & 1.19 \\
%2964& 1.28 &4254 & 1.28 & & 
%1042 & 1.47 &
%3423 & 1.28 &4487 & 1.28 \\
\hline
\end{tabular}}
\caption[\small Values for the colour $V - I$ adopted in the calibration of the optical images from HST.]
{\small Values for the colour $V - I$ adopted in the calibration of the optical images from HST.}
\label{centralcomp}
\label{tab6ch4}
\end{table} 
\indent Four of the galaxies for which we perform our optical analysis (NGC\,1042, 3423, 4030 and 4102) belong also to the sample
studied by \citet{pohlen}, who present SDSS $g'$ and $r'$ profiles (roughly corresponding to $B$ and $R$ bands) for 90 almost face-on 
late-type galaxies. The authors kindly provided us their profiles for the galaxies in common. In their profiles we recognize all the
features that we see in ours, at the same radii, out to $\approx$ 150 - 200\arcsec\, (depending on the galaxy): the profiles are
completely consistent, with the caveat that they are  measured in different bands.\\
\begin{figure*}\centering
{\includegraphics[width=\textwidth]{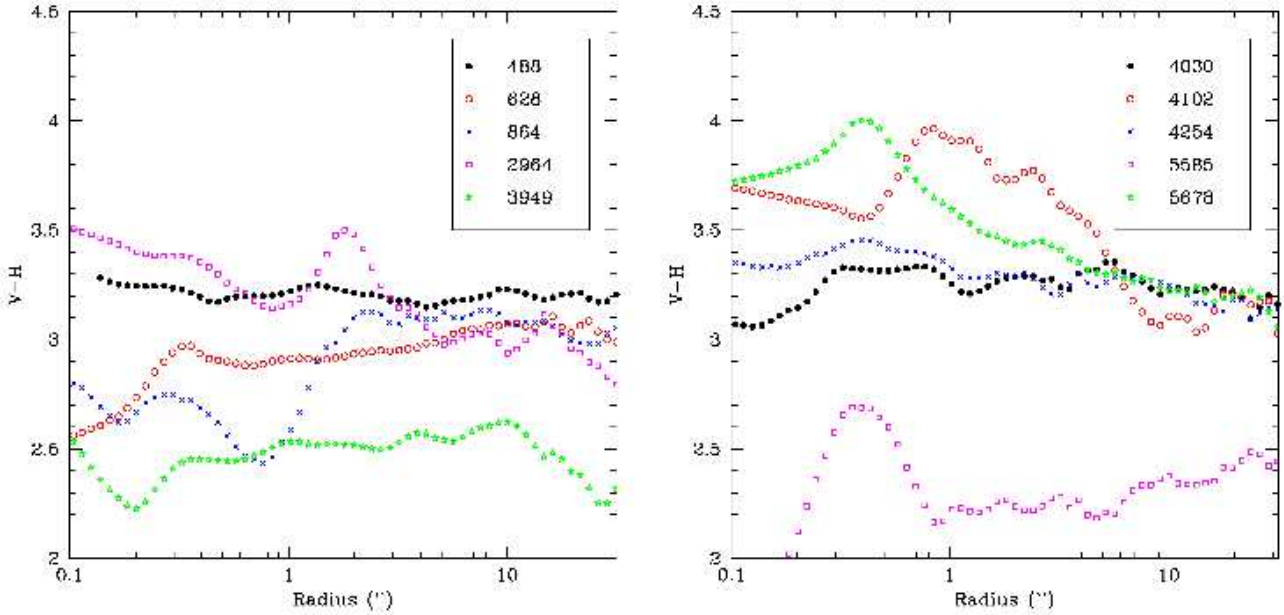}}
\caption{V - H colour profiles for the 10 galaxies with available optical and
NIR HST images}
\label{fig20ch4}
\end{figure*}
\begin{figure*}\centering
{\includegraphics[width=8truecm]{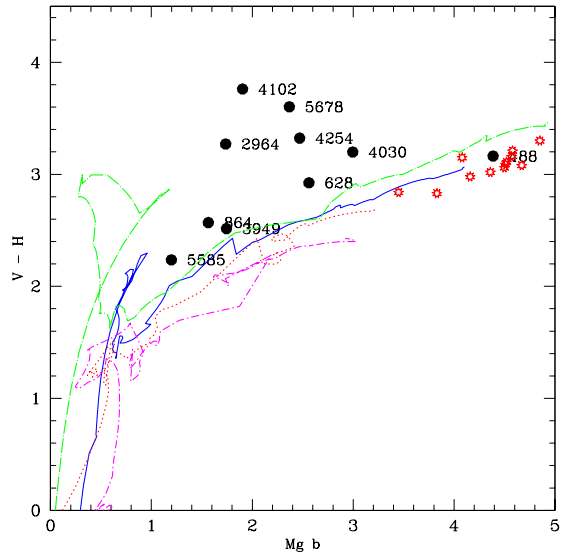}}
\caption{$V-H$ colour vs. Mg~b line strength in central apertures of 2.4$''$ for the 10 galaxies with NICMOS
F160W and F606W HST data (black filled symbols). In red (asteriscs) elliptical and S0 galaxies are shown
from Kuntschner et al. (2006), with $V-K$ measurements from Frogel et al. (1978). Lines represent SSP models
of Bruzual \& Charlot (2003) for various metallicities: green (long dashed-dotted) indicates Z=0.05, blue,
full lines indicate Z=0.02, red dotted line Z=0.008 and magenta short-dashed-dotted lines Z=0.004. 
}
\label{fig19ch4}
\end{figure*}
\setcounter{subfigure}{1}
\begin{figure*}\centering
{\includegraphics[height=21truecm]{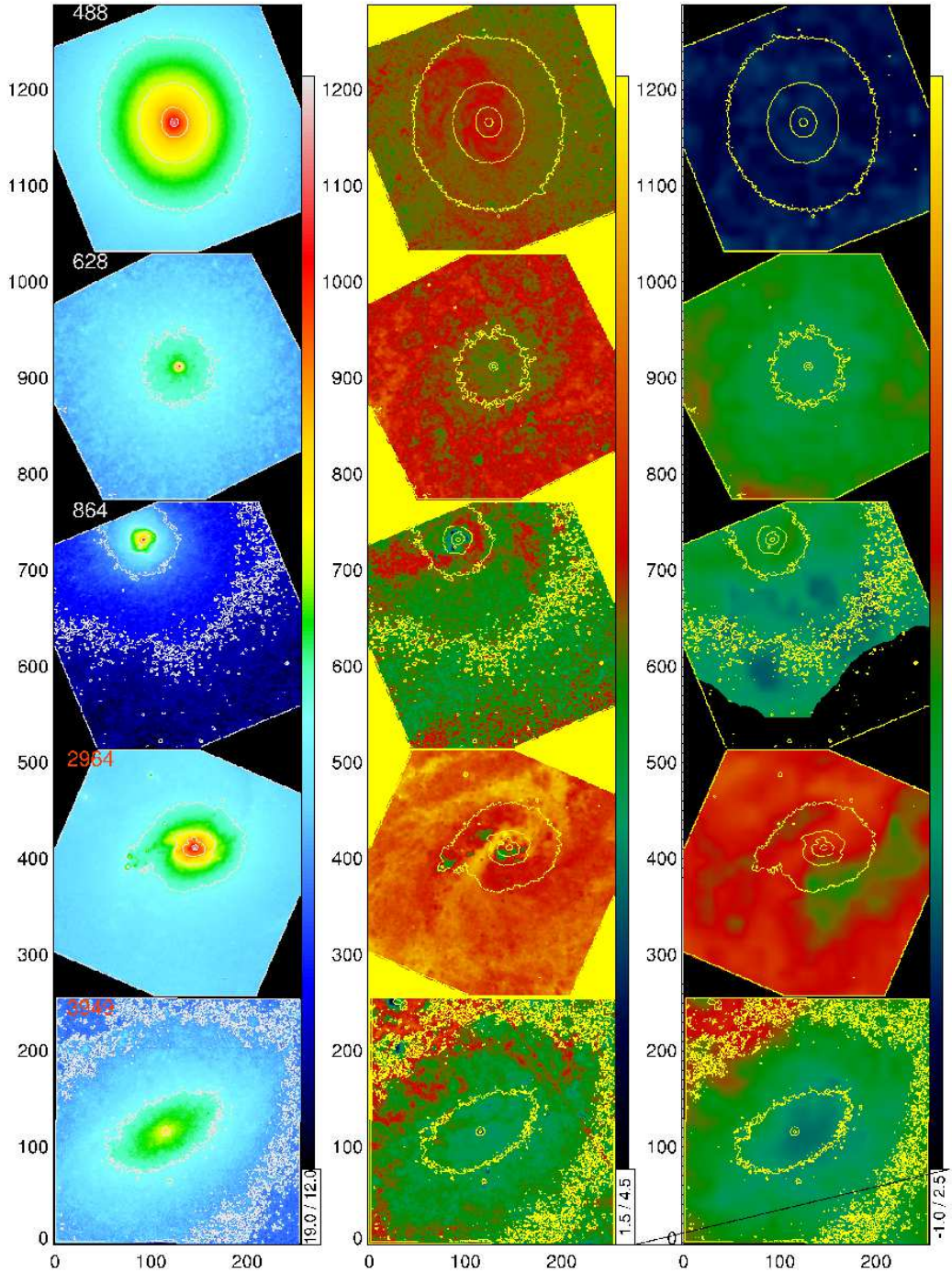}}
\caption{(a): Extinction distribution for the 10 galaxies with space-based photometry in
F606W and F160W. Left: surface brightness distribution in H. Middle: calibrated
V-H colour maps. Right: extinction maps from $V-H$ and Mg~b. The ranges of the
plots are shown in the lower right of each column of plots.}
\label{fig22ch4}
\end{figure*}
\addtocounter{figure}{-1}
\addtocounter{subfigure}{1}
\begin{figure*}\centering
{\includegraphics[height=21truecm]{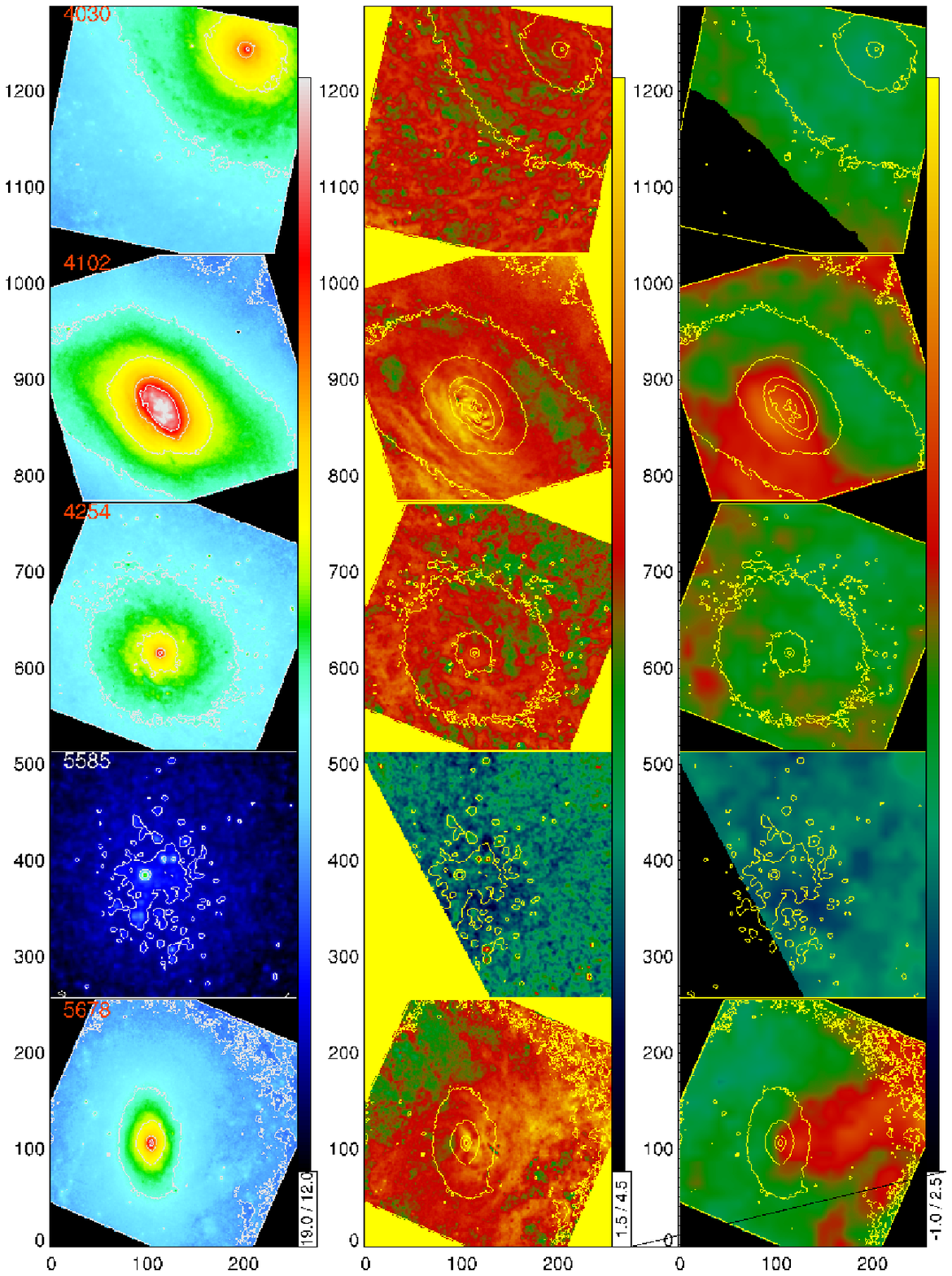}}
\caption{(b): Extinction distribution for the 10 galaxies with space-based photometry in
F606W and F160W. Left: surface brightness distribution in H. Middle: calibrated
V-H colour maps. Right: extinction maps from $V-H$ and Mg~b.}
\label{fig23ch4}
\end{figure*}

\indent From the thus obtained $V$ and $H$ profiles we determined radial $V -
H$ colour profiles, that we present in Fig. \ref{fig19ch4}. Here we only
present the inner 20$''$, since real near-infrared imaging is only available in
this region. Further out, the 2MASS images are not deep enough. We find that
the profiles have large dispersions in colour, and a wide variety in slope.
Some galaxies become redder, some bluer as a function of radius. When comparing
them to elliptical galaxies (e.g. Peletier, Valentijn \& Jameson 1990) the
profiles are much less smooth, and the gradients much larger. The colour maps
are very instructive in telling us what is happening. Fig. \ref{fig22ch4}a and
\ref{fig23ch4}b show the calibrated $V-H$ colour maps of the inner 20$''$
$\times$ 20$''$ of the galaxies. They have been displayed on the same scale, to
make it easier to make the comparison from galaxy to galaxy, and to compare the
internal gradients with the colour differences from galaxy to galaxy. The
colour maps show dust lanes in all galaxies except for NGC~5585, spiral arms
with younger stars, inner spirals, nuclear clusters, etc. In most cases it is
not easy to find out where the galaxy center is, when just looking at the
colour map. 

\subsection{Extinction in the centers of late-type spiral galaxies} From a $V-H$ map alone it is very
difficult to determine the amount of extinction E$_{V-H}$. The problems is that the intrinsic colour of the
stars is unknown, so that the uncertainty in the colour of the stars is reflected in the uncertainty in the
amount of extinction. Having two colours does not help very much, since the effects of reddening in
colour-colour diagrams is almost parallel to the effect due to changing metallicity or age (e.g. Kuchinski
et al. 1998), and since the distribution of the dust changes the relative extinction in the different bands.
Determining the amount of extinction in a specific galaxy is therefore done assuming a certain colour for
the stellar populations (see e.g. Knapen et al. 1995), using the Balmer decrement for ionised gas (which
usually gives the extinction on the lineof sight to an HII region, but not in the whole galaxy), or using
statistical methods. For example, from the distribution of colour as a function of inclination one can
derive the amount of extinction (e.g. Tully et al. 1998, Giovanelli et al. 1995, Peletier et al. 1995). The
SAURON dataset offers the nice advantage that absorption line indices are available in the inner 40$''$
$\times$ 30$''$. Since to first order the Mg~b index in a galaxy gives the same information as the $V-H$
colour (which is the metallicity for an old galaxy, see Fig. \ref{fig19ch4}), and since Mg~b is affected
very little by extinction (MacArthur 2005), one can use the $V-H$ and Mg~b together to determine the amount
of extinction at every position where both are available. Here we use the central Mg~b measurements of Ganda
et al. (2007) to calculate the extinction in the central aperture with diameter 2.4$''$. In
Fig.~\ref{fig19ch4} these central values are shown. The figure shows that the models with various
metallicities almost fall on top of each other. This means that one can measure the extinction by measuring
the distance in $V-H$ to a line of models with a given reference metallicity. Since most of the galaxies
have luminosities similar to the Milky Way, we take the solar models as a reference. For NGC~5585, which is
fainter, it is probably more appropriate  to take a lower metallicity. The uncertainty in the amount of
extinction is then given by the uncertainty in the $V-H$ of the stellar populations in the galaxy that make
up the Mg~b. From the dispersion  between the SSP models we estimate that this uncertainty is about 0.2 mag
or less, even if the stellar populations consist of a mix of ages, since more complicated models are always
linear combinations of SSPs. The obtained E$_{V-H}$ values are converted to A$_V$ using the Galactic
extinction law (Rieke \& Lebofsky 1985), dividing them by 0.825. Here the assumption is made that the
galaxies are not optically thick in $H$. Given the extinction values that are obtained, this assumption is
easily satisfied. Table 4 shows the extinction values we derived. We find an average E$_{V-H}$ of
0.56 mag, or a lower limit to the average extinction A$_V$=0.68 mag. In Fig.~\ref{fig22ch4} the central
extinction maps, determined in this way, at the SAURON resolution, are shown.

\begin{table}\centering
\small{\begin{tabular}{@{}l l l  l | l l l l}
\hline \hline
NGC &  $V-H$ & E$_{V-H}$ & A$_V$& NGC & $V-H$ & E$_{V-H}$ & A$_V$ \\
\hline
488 & 3.16 & 0.00 & 0.00 & 4030 & 3.20 & 0.47 & 0.57 \\
628 & 2.92 & 0.33 & 0.40 & 4102 & 3.76 & 1.44 & 1.75 \\
864 & 2.57 & 0.35 & 0.42 & 4254 & 3.32 & 0.74 & 0.90 \\
2964& 3.27 & 0.90 & 1.09 & 5585 & 2.24 & 0.22 & 0.27 \\
3949& 2.52 & 0.14 & 0.17 & 5678 & 3.60 & 1.05 & 1.27 \\
\hline
\end{tabular}}
\caption[]
{\small $V-H$ and central extinction in an aperture of diameter 2.4$''$ derived from $V-H$ and Mg~b (see Fig. \ref{fig19ch4}).}
\label{centralcomp}
\label{tab6ch4}
\end{table} 

\begin{figure*}\centering
{\includegraphics{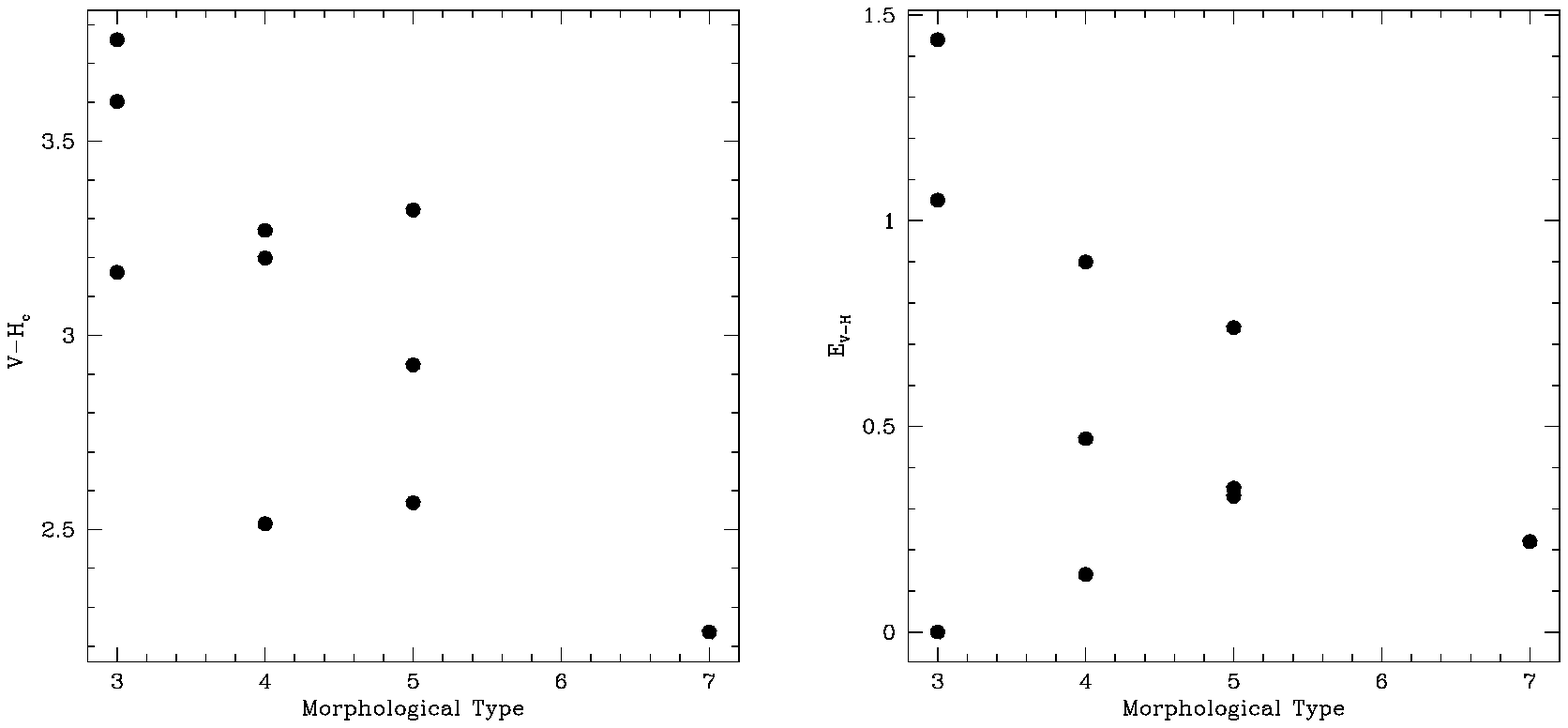}}
\caption{Left: Central V-H colour as a function of morphological type. 
Right: Central extinction E$_{V-H}$ as a function of morphological type.}
\label{fig60ch4}
\end{figure*}

Do we learn anything about the stellar populations from the $V-H$ colour? Figure \ref{fig19ch4} shows that we don't. For some galaxies
$V-H$ is redder than for the most massive giant ellipticals. Since the line indices of all objects are weaker than those of massive
giant ellipticals, we have a strong indication that it is dust extinction, and not large numbers of very red stars, that causes the red
$V-H$ colour. This argument is made stronger by the fact that for highly inclined galaxies optical-infrared colours in regions outside
the dustlane always are bluer than those of giant ellipticals (Peletier \& de Grijs 1998). The issue of the very red optical-infrared
colours was already discussed in 1985 by Frogel, who found galaxies with $V-K$ colours that were considerably redder than those of the
brightest giant ellipticals. He found also that half of his sample of Sc galaxies had $U-V$ colours that were bluer than the giant
ellipticals. Although he could not uniquely determine the extinction, he used the following experiment. If all galaxies had an intrinsic
stellar population colour of $J-H$=0.65, he obtained extinction values between 0 and 1.64 mag in A$_V$. The optical-infrared colours
would not be redder any more than those of giant ellipticals, and also reasonable $U-V$ colours would be obtained. Independently,
Turnrose (1976) had derived internal extinction by comparing simple stellar population models with observed optical stellar energy
distributions, and had found similar extinction values. We have one galaxy in common with Frogel (1985), NGC 628, for which $V-H$ agrees
within 0.1 mag.

Since these are all close to face-on galaxies, the measured extinctions are not very different from what one would see if they were
face-on. The extinction correction to face-on is given by, $$\Delta \mu_V ~=~ \gamma~log(a/b) $$ where a/b is the axis ratio of the
galaxy, and $\gamma$ a transparency index (Boselli \& Gavazzi 1994). For this sample $log (a/b)$ ranges from 0 to 0.26, while values for
$\gamma$ should be between 0.4 and 0.7 (Boselli \& Gavazzi 1994), so that the face-on corrected A$_V$ is between 0 and 0.2 mag less than the
value given in Table \ref{tab6ch4}, sometimes still around or above 1 mag. This result is compatible with Tully et al. (1998), Peletier
et al. (1995) and Giovanelli et al. (1995), who used a statistical method, deriving the extinction from the relation of colour, or
surface brightness, as a function of inclination. 

We now look at extinction as a function of morphological type. This is shown in Fig.~\ref{fig60ch4}. The relation is not very clear, but
it can be understood when one looks at the outliers. The galaxy of morphological type 7 (Sd, NGC 5585) is by far the bluest, and the
smallest and has little extinction. This is in agreement with Tully et al. (1998) who find that galaxies with M$_B$=-18 have almost no
extinction. At the other end, at morphological type 3=Sb, we find a galaxy with very little extinction, NGC 488, and two others with a
lot of extinction. This can be understood by looking at Fig.~12 of Terndrup et al. (1994), 
where one can see that galaxies with type
3-5 have much more extinction than galaxies of earlier types. Peletier \& Balcells (1996) 
confirm this, showing also that the extinction in galaxies of type 7 and larger is again much
smaller than in galaxies of type 3-6. It looks as if NGC~488 is more similar to an Sab galaxy than an
Sb in its inner regions. One should note that the A$_V$ values that we have derived here are still lower limits. For example, by adding
very young stellar populations mixed with dust, one can still make the color of a galaxy bluer. For that reason, the conclusion that the
center of NGC~3949 is almost dust free is most likely wrong looking at the structure in the colour map.
Our derived extinction values for these late-type spirals are consistent with Tuffs et al. (2004), who
use the models of Popescu et al. (2000) to determine the attenuation of dust in spirals using a thin
and a normal disk with extinction, and a bulge without dust. We should note that this models has a
very large optical depth ($\tau_{B,cen} = 3.8$) so that, according to the model, we basically only see
the bulge light in front of the disk, as a result of which the integrated reddening is not so high,
and the colours reasonable. With these data we are not able to verify whether this extinction model is
realistic for late-type spirals.

Does the finding that many of our late-type galaxies have a considerable amount of extinction in
their centre have other consequences? Martini et al. (2003), and Carollo et al. (2002) also show
optical-infrared maps of similar galaxies, some of which are in common with us. They however do not
show any calibrated color maps. To test the effects of extinction on the structural parameters, we
also calculated the same parameters as presented in Section 5 in the $V$-band. In general, the
differences were small, something which could have been expected from the fact that for these
galaxies the colour profiles are generally reasonably flat. Since there are many galaxies with large
colour gradients (see e.g. Martini et al. 2003 and Carollo et al. 2002) we think it is preferable to
determine these parameters in the NIR only.
 
Peletier \& Balcells (1996) show that colours of bulges and disks of spirals, measured in relatively
dust free regions, are very similar, implying that stellar populations in bulges and inner disks are 
similar. Terndrup et al. (1994) find the same effect, but with more scatter in the colours, and with
a sample that includes redder galaxies. As we do here, they claim that these galaxies are red
because of internal extinction. The fact that bulge and disk colours, even for these red galaxies,
are similar shows that similar amounts of extinction are present in both components. If galaxies are
seen close to face-on, i.e. when most of the light comes from regions close to the symmetry plane of
the galaxy, the colors of bulge and inner disk region both are affected by similar amounts of
extinction. Images of edge-on galaxies show that the extinction is mainly situated in the plane.

Carollo et al. (2001) present a study of $V-H$ colours in a sample of bulges, also from HST imaging.
To convert from her ${V-H}_{AB}$ colours to the Johnson system we use here we need to add 1.45 mag.
By masking out patches of dust and star formation regions, they obtain $V-H$ colours that are blue
enough that they can be explained by stellar population models, for both types of bulges discussed.
However, the nuclei have often much redder colours, see their Fig.~3, that need considerable
extinction. The problem with selecting {\it clean} regions by hand is that this method is subjective
and probably not reproducible, giving results that are biased towards blue colours. The same is done
in Carollo et al. (2007), where only one galaxy is found with optical-infrared colours redder than
those of ellipticals and early-type bulges.

Dust can have important consequences for the interpretation of mass profiles from rotation 
curves and optical photometry. Without correcting for extinction, which is often impossible,
the resulting stellar M/L ratios will be too large, and show a much larger scatter than is due
purely by stellar populations. Since the dust extinction is largest in the central regions, and
HI observations often have a limited angular resolution, the effect is most important when
interpreting optical rotation curves. Palunas \& Williams (2000) find that if they correct the
surface brightness by $A_{I,int} = 1.0{\mathrm log}({a \over b})$ the resulting M/L ratios do
not vary any more as a function of inclination. In the $V$-band the correction should be larger
by a factor $\sim$1.5. Such corrections in general are not very meaningful, since extinction
varies as a function of radius. If one wants to obtain accuracte M/L ratios in the centers of
galaxies, one either has to work in the NIR only, or make detailed extinction corrections.

\section[Bulge and disk parameters]{BULGE AND disk PARAMETERS}\label{correlationsecch44}

Here we study the structural parameters determined in the previous
Section together with data from the literature. 
%Global scaling relations are extremely important, since they allow us to test the formation
%models of the various galaxy components; for the bulge component, comparisons
%with the relations for ellipticals can help understanding if and how much bulges
%are similar to elliptical galaxies, shedding some light on their nature.

%In the following,
We  quantitatively investigate the 
structural parameters obtained from the bulge-disk decomposition
as a function of various fundamental galaxy parameters, such as morphological
type, central velocity dispersion, and luminosity, to obtain a better
understanding of the formation of these galaxies. 
%Here we show a few noteworthy relations. More relations are shown in Appendix B. There, for example, we look at relation between bulge and disk parameters. 
Only the most noteworthy relations are shown here; additional relations, including those 
between disk and bulge parameters, are shown in Appendix B.  

We concentrate, in particular, on how late-type spiral galaxies differ from earlier types. 
We do this by
comparing our galaxies with the large literature compilation of Graham \& Worley (2008),
and with the recent sample of Fisher \& Drory (2008). Graham \& Worley present a
compilation of about 10 samples, mostly in the near-infrared. Fisher \& Drory analyse a
sample of HST data, mostly in the optical. For the relations with central velocity
dispersion we compare with the sample of early-type galaxies of Balcells et al. (2007b),
hereinafter B07. For the literature samples, we took the numerical values from
the tables available in the published papers, adjusting the scale lengths and
effective radii in order to homogenise the adopted value for the Hubble constant
to $H_{0}$ = 70 km s$^{-1}$ Mpc$^{-1}$ (for consistency with what done by
\citealt{ganda} 
and with the luminosities
calculated in Section \ref{lumsec}) . The total magnitudes (of bulge and disk) of
\citet{balcellsglobal} were converted from $K$ to $H$ using the  typical colour for disk
galaxies $H - K = 0.21$ mag (Frogel et al. 1978). For most of the relations that we will
investigate and show graphically, we list in Table \ref{tab3ch4} the linear Pearson
correlation coefficient $c$, calculated for the galaxies of our sample only. For the
strongest correlations, with $\vert c \vert \ge 0.5$, we performed an outliers-resistant
linear fit, assuming a dependence of the form $y = a + b \times x$, $y$ and $x$ being the
variables involved and $a$ and $b$ the fit parameters. Those fits, for which the
parameters are given in Table  \ref{tab4ch4}, are discussed in detail below.

\subsection{Trends among the structural parameters}

\begin{figure*}\begin{center}
{\includegraphics[width=17truecm]{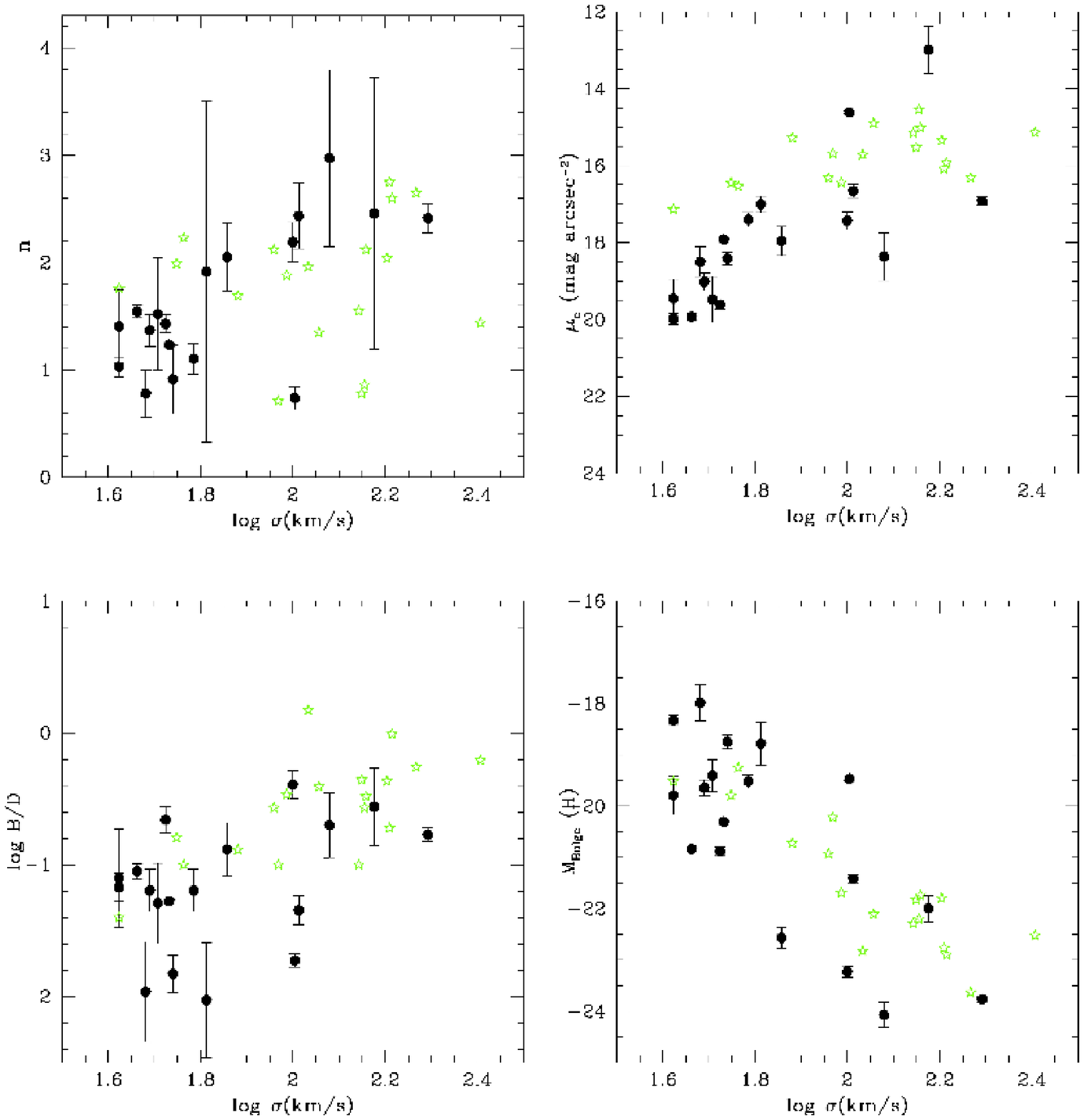}}
\caption{Structural parameters as a function of central stellar velocity 
dispersion for our late-type sample (black filled circles), compared to the 
sample of early-type spirals of \citealt{balcellsglobal} (green asteriscs). 
The velocity
dispersions for our sample are taken from Table 1 in \citet{ganda}. As for the
morphological T-type, we have artificially put all the galaxies earlier than T =
0 in the sample of \citealt{balcellsglobal}
to T = 0.}
\label{fig10ch4a}
\end{center}\end{figure*}

In Fig.~\ref{fig10ch4a} we show a number of structural parameter as a function of
stellar velocity dispersion. Here one should note that due to the paucity of velocity
dispersions in the literature we can only compare 2 samples of
about 20 galaxies, of resp. early and late type spirals. There are some interesting
things to note. Several papers, e.g. B07, Fisher \& Drory (2008) and Graham \& Worley 
(2008), show that the S\'ersic index $n$ correlates reasonably well with
bulge luminosity, with galaxy type and bulge to disk ratio, but with considerable
scatter. Fig.~\ref{fig10ch4a}a shows that the same holds for the central velocity
dispersion $\sigma$. There are late-type spirals with velocity dispersions larger than
100 km/s, that show $n$-values considerably larger than 1. If all bulges of late-type
spiral galaxies would be disks (also called pseudo-bulges), their $n$-values would be
close to 1. This is however not the case, and it appears that the central velocity
dispersion of the galaxy is determining the shape of the surface
brightness profile of the bulge, rather than the galaxy type, which is determined more by the
total star formation or the bulge to disk ratio, Consistent with this picture is the relation between
bulge luminosity and central velocity dispersion. Already B07 found a tight relation
between these 2 quantities for early-type spiral galaxies. Here we see that late-type
spirals follow the same relation. So, even though the surface brightness of bulges of
late-type spirals is lower than of early-type spirals, and the B/D ratio for a given
velocity dispersion is lower as well (Fig.~\ref{fig10ch4a}bc), the
bulge luminosity - $\sigma$ relation is the same for both samples of galaxies. Since
there is a strong correlation between the bulge luminosity, or central surface
brightness, or S\'ersic index $n$ (Graham \& Driver 2007), and black hole mass, 
a relation which also holds for the luminosity and black hole mass of elliptical galaxies and
S0's, it is tempting to assume that bulges of late-type spirals also lie on this
relation, and that the formation of those bulges is closely linked to the black holes in
late-type spirals. 
If a central black hole is associated with every
bulge, it would also explain why the tight black hole mass vs. central velocity
dispersion relation (Ferrarese \& Merritt 2000, Gebhardt et al. 2000) can 
be reproduced with bulge
luminosity replacing central velocity dispersion, but not with disk or total
galaxy luminosity. The lack of correlation between the sizes of bulges and disk
with central velocity dispersion for late type spirals might also be 
a result of the fact that bulges of late-type spirals are less concentrated 
objects with $n~\sim~ 1$, which do not follow the homology of elliptical
galaxies and elliptical-like bulges. 

We also like to point out that it appears that bulges of early and late-type galaxies
show systematic differences: on the average early type spirals with low central
velocity dispersion have much more concentrated bulges and disks: not only are 
disks and bulges smaller, but also the surface brightness of both components is higher.

\subsection{Trends with Stellar Populations}
This sample is unique in the sense that its central stellar populations have
been studied using stellar absorption lines, which are almost independent of
extinction, by Ganda et al. (2007), so that we can investigate whether the
stellar populations depend on the structural parameters discussed above. 
In other words, are the age and star formation history of a
galaxy related to the way it is built, and viceversa? We try to address this
issue by plotting the results of our bulge-disk decomposition against population
parameters. A few noteworthy relations are discussed here. The rest of the relations are
discussed in Appendix B.  

Fig. \ref{fig6n} shows the S\'ersic parameter $n$ and the disk central surface 
brightness and as a function of the central
H$\beta$ absorption, a measure of the age of the galaxy, and the 
star formation time-scale $\tau$. 
The stellar population parameters refer to a central aperture of radius 1\farcs5.
One sees that $n$ increases of decreasing H$\beta$. This implies that for objects with
relatively young stellar populations, bulges are exponential, and that old bulges are
more similar to objects with a de Vaucouleurs distribution. This would be the case if the
objects in which recently stars have formed are all disk-like, i.e. pseudobulges, and
that old bulges are more spheroidal-like. The latter statement is very well applicable to
bulges of early-type galaxies (e.g. Kormendy \& Kennicutt 2004, Fisher \& Drory 2008),
but is not so obvious for late-type spirals. It implies that some late-type spirals have
old bulges, that do not follow exponential distributions. Maybe our Milky Way galaxy has
a bulge similar to those. 
We also see that $n$ decreases for increasing star formation time scale $\tau$. Late-type
spiral disks, like the disk of our Milky Way, have long star formation time scales. As a
result, parameters such as H$\beta$, which are strongly influenced by the last burst of
star formation, generally show large values, indicative of the presence of young stellar
populations. We find that galaxies with small star formation time scales, i.e.
elliptical-like objects, have high $n$-values.
Interesting are also the correlations with disk scale length. If there are signs of
younger populations in the center, or if the star formation time scale is long, then the
disks are generally smaller. This effect is probably due to the fact that 
more massive galaxies
generally are of earlier-type, having a shorter star formation time scale, and show older
bulges.

\subsection{Trends between structural parameters}
Coming to the relations among the structural parameters themselves, in Fig.
\ref{fig14ch4} we plot the size and surface brightness of the bulges and disks
of our galaxies against each other, and add the samples of GW and Fisher \&
Drory (2008) on the right: in the left
panel we plot the central surface brightness of the disk 
against the effective
surface brightness of the bulge; in the right one, 
disk scale length against bulge effective
radius. 
We see a strong trend that galaxies with brighter disks harbour brighter bulges, 
and a tendency  for galaxies with larger disks to have larger bulges (see
before). The figure shows conclusively that bulge and disk are related, and that
e.g. high surface brightness bulges do not co-exist with low surface brightness
disks and vice-versa. The figure suggests that disks do not form without
affecting the bulge. If a disk is formed in a galaxy which already contains a
bulge, its size and surface brightness is either determined by the mass
distribution of the bulge, or the bulge adapts itself, implying that stars are
being formed in this process. The strong correlation between $\mu_{e,Bulge}$ and
$\mu_{0,Disk}$ seems to support the process of the formation of bulges through 
secular evolution of disks.

%To conclude this
%section, in Fig. \ref{fig15ch4} we look at the trends of bulge and disc
%parameter with $n$: bulge effective surface brightness, disk central surface
%brightness, bulge effective radius, disk scale length, $r_e$/h ratio and B/D
%luminosity ratio, for several samples. We omit the sample of \citet{jong96},
%since there $n$ is fixed to 1. We find a global trend for r$_{e}$ and also for
%B/D to increase when $n$ increases, but no tight correlation; in general in this
%figure we observe a significant dispersion among the samples.

\begin{table}
\centering
\tabcolsep=1.3mm
\begin{tabular}{@{}l l c | l l c}
\hline \hline
x & y & $c$ & x& y & $c$\\
(1) & (2) & (3) &(4) & (5) & (6)\\
\hline
T & $n$ & -0.599 & $\log(\sigma)$ & $n$ &  0.641\\
T & $\mu_{e,b}$ & 0.769 & $\log(\sigma)$ & $\mu_{e,b}$ &-0.742\\
T & $\log (r_e)$ &  0.155 & $\log(\sigma)$ &$\log (r_e)$ & 0.028\\
T & $\mu_{0,d}$ & 0.843 & $\log(\sigma)$ & $\mu_{0,d}$ & -0.720\\
T & $\log (\mathrm{h})$ &0.045& $\log(\sigma)$ & $\log (\mathrm{h})$ &0.113\\
T & $r_e$/h &0.150 & $\log(\sigma)$ & $r_e$/h &-0.065\\
T & $\log$ (B/D) & -0.256 & $\log(\sigma)$ &$\log$ (B/D) & 0.416\\
\hline
M$_d$ & $n$ & -0.679 & M$_b$ & $n$ & -0.705\\
M$_d$ &$\mu_{e,b}$ &0.394 & M$_b$ &$\mu_{e,b}$ &0.276\\
M$_d$ &$\log (r_e)$ &-0.297 & M$_b$ &$\log (r_e)$ &-0.567\\
M$_d$ &$\mu_{0,d}$ &0.490 & M$_b$ &$\mu_{0,d}$ &0.416\\
M$_d$ &$\log (\mathrm{h})$ &-0.555 & M$_b$ &$\log (\mathrm{h})$ & -0.414\\
M$_d$ &$r_e$/h &-0.022&M$_b$ &$r_e$/h &-0.483\\
M$_d$ &$\log$ (B/D) &-0.260&M$_b$ &$\log$ (B/D) &-0.794\\
\hline
H$\beta$ & $n$ & -0.681 &H$\beta$ & $\log (\mathrm{h})$ & -0.532\\
$\Delta$Mg{\textit{b}}$'$ & $\log (r_e)$& 0.724& $\Delta$Mg{\textit{b}}$'$ & $\log (\mathrm{h})$ & 0.759\\
$\log(\tau)$ & $n$ &  -0.678 & $\log(\tau)$ & $\log (\mathrm{h})$ & -0.633\\
\hline
M$_{Bulge}$ & $\log(\sigma)$ &-0.756 & M$_{Bulge}$ & T & 0.630 \\
\hline 
$\mu_{e,b}$ & $\mu_{0,d}$ & 0.866 & $\log (r_e)$ & $\log (\mathrm{h})$& 0.693\\
\hline
$n$ & $\mu_{0,d}$ & -0.430 & $n$ & $\mu_{e,b}$& -0.338\\
$n$ & $\log (\mathrm{h})$ & 0.285 & $n$ & $\log (r_e)$ & 0.231\\
$n$ & $r_e$/h & 0.143 & $n$ & $\log$ (B/D) & 0.440\\
\hline
\end{tabular}
\caption{Correlation coefficient $c$ for the parameter-parameter relations, for
our sample only. We report the correlation coefficient for all the relations of
Figs. \ref{fig10ch4aa} and b, and \ref{fig11ch4a} and b; 
for those of Figs. \ref{fig12ch4} we only
list those with $\vert c \vert \ge 0.5$. Columns (1), (2) and (4), (5):
variables; columns (3) and (6): correlation coefficient. For the relations
involving the T-type, all the individual points are considered and not the
average values within a type (as done instead in Fig. \ref{fig10ch4aa} and b).}
\label{tab3ch4}
\end{table}

\begin{table}
\vspace{0.3cm}
\centering
\tabcolsep=1.3mm
\scriptsize{\begin{tabular}{@{}l l | c c | l l | c c }
\hline \hline
x & y & $a$ & b & x & y & $a$ & b \\
(1) & (2) & (3) &(4) & (5) & (6) & (7) & (8)\\
\hline
T & $n$ &  3.104  &  -0.287& $\log(\sigma)$ & $n$ &  -2.654  &    2.332\\
T & $\mu_{e,b}$ & 14.644   &  0.688 & $\log(\sigma)$ & $\mu_{e,b}$ &30.036  &   -6.628\\
T & $\mu_{0,d}$ & 14.346  &    0.623 & $\log(\sigma)$ & $\mu_{0,d}$ & 24.903  &    -4.017\\
\hline
M$_d$ & $n$ & -7.970  &   -0.410 & M$_b$ & $n$ & -4.941 &   -0.316\\
M$_d$ &$\log (\mathrm{h})$ & -2.045 &   -0.108& M$_b$ &$\log (re)$ & -2.936   &  -0.133\\
& & & & M$_b$ &$\log$ (B/D) &-5.292  &   -0.200\\
\hline
H$\beta$ & $n$ &  4.329  &   -0.876&H$\beta$ & $\log (\mathrm{h})$ & 1.157   &  -0.221\\
$\Delta$Mg{\textit{b}}$'$ & $\log (r_e)$& 0.353    &   14.966& $\Delta$Mg{\textit{b}}$'$ & $\log (\mathrm{h})$ & 0.807   &    8.358\\
$\log(\tau)$ & $n$ &  2.638  &   -1.590 & $\log(\tau)$ & $\log (\mathrm{h})$ & 0.782 &   -0.481\\
\hline
M$_{Bulge}$ & $\log(\sigma)$ & 0.190  &  -0.081 & M$_{Bulge}$ & T & 15.442  & 
0.506 \\
\hline 
$\mu_{e,b}$ & $\mu_{0,d}$ & 8.092  &    0.527 & $\log (r_e)$ & $\log (\mathrm{h})$& 0.566    &  0.373\\
\hline
\end{tabular}}
\caption{Parameters of the linear best-fit for the relations with $\vert c \vert
\ge 0.5$ in Table \ref{tab3ch4}. The relation $y = a + b \times x$, with $y$ and
$x$ variables (columns (1), (2), (5), (6)) and $a$ and $b$ fit parameters
(columns (3), (4), (7), (8)) is fitted to the data for our sample only. For the
relations involving the T-type, all the individual points are considered and not
the average values for the same type (as done instead in Fig.~\ref{fig10ch4aa}
and b).}
\label{tab4ch4}
\end{table} 

\begin{figure*}\begin{center}
{\includegraphics[width=\textwidth]{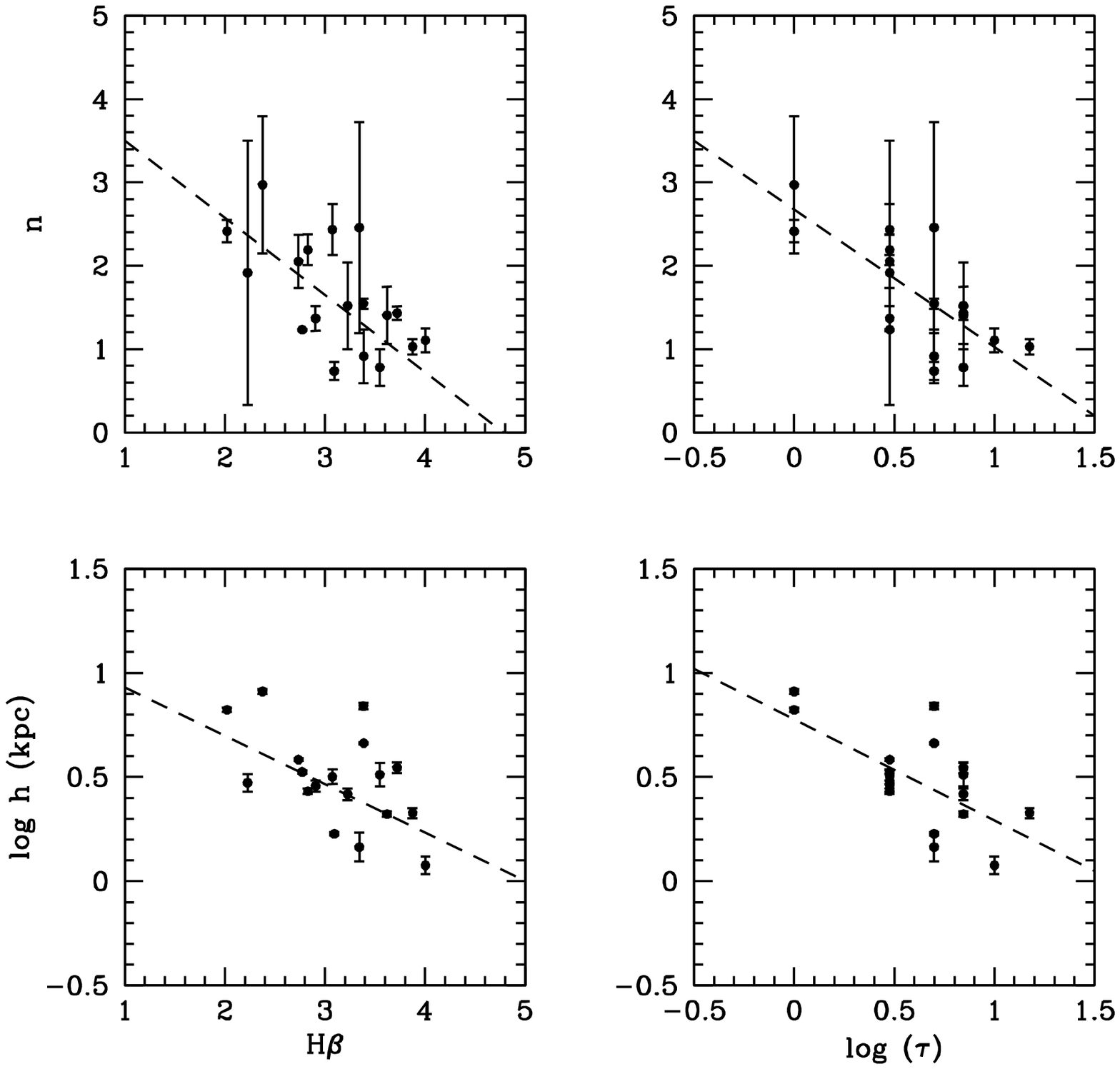}}
\caption{Correlations between the structural parameters $n$ and disk scale
length $h$ as a function of stellar population
parameters H$\beta$ (in \AA) and star formation timescale $\tau$ (in Gyr). 
The dashed lines overplotted represent a linear
fit to our data.}
\label{fig6n}
\end{center}\end{figure*}

\renewcommand{\thefigure}{\arabic{figure}}
\setcounter{subfigure}{1}
\begin{figure*} \centering
{\includegraphics[width=\textwidth]{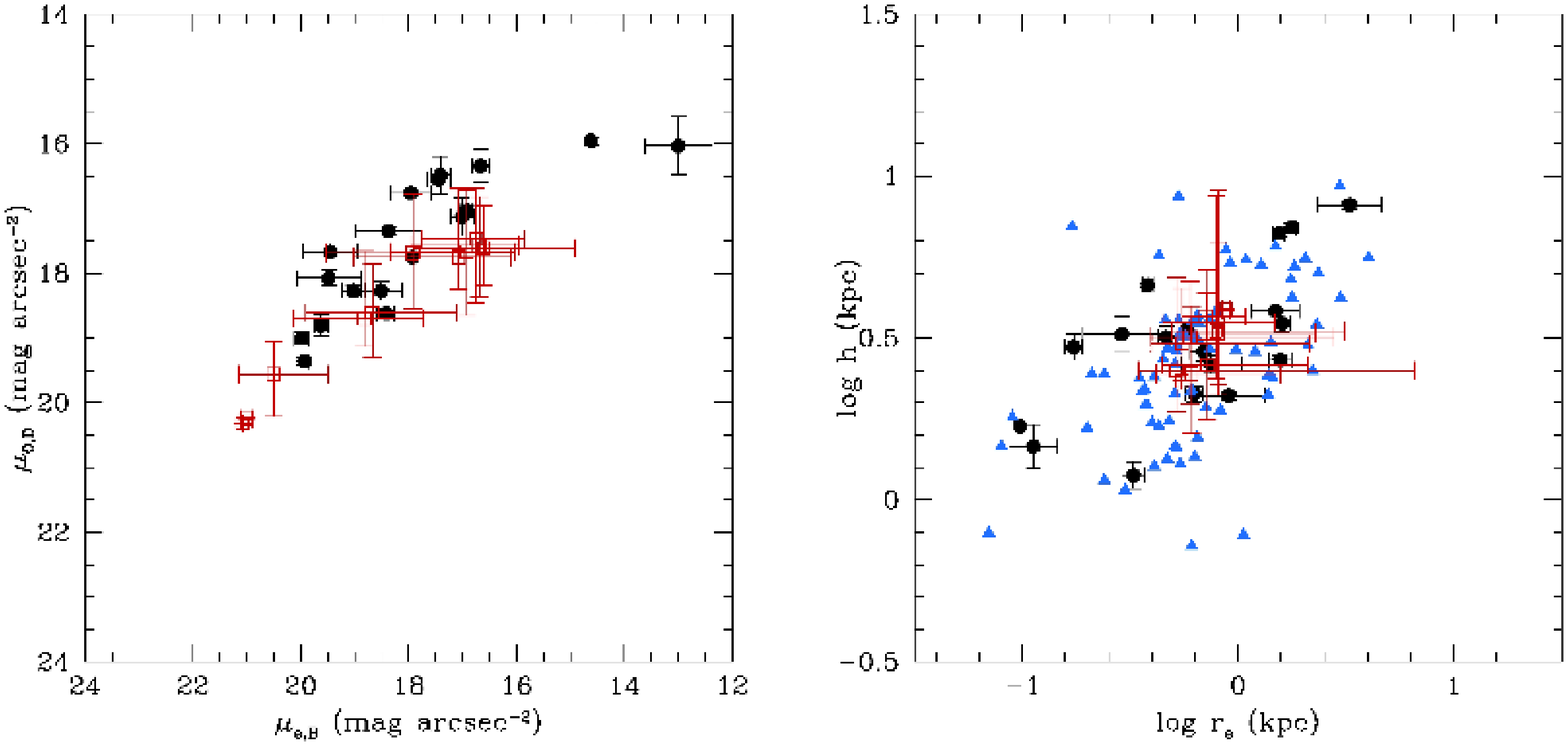}}
\caption{Bulge against disk parameters: {\textit{left panel}}: disk central
brightness $\mu_{0,Disk}$ against bulge effective surface brightness $\mu_{e,Bulge}$
measured in mag arcsec$^{-2}$; {\textit{right panel}}: disk scale length $h$
against bulge effective radius $r_{e}$, both measured in kpc. Filled dots are the 
galaxies of this paper. Blue triangles are galaxies from Fisher \& Drory (2008), 
while the open red squares indicates the compilation of Graham \& Worley (2008).
}
\label{fig14ch4}
\end{figure*}

\section[Central components]{CENTRAL COMPONENTS}\label{centralcompsecch4}
\begin{center}
\begin{figure}\begin{center}
{\includegraphics[width=0.49\textwidth]{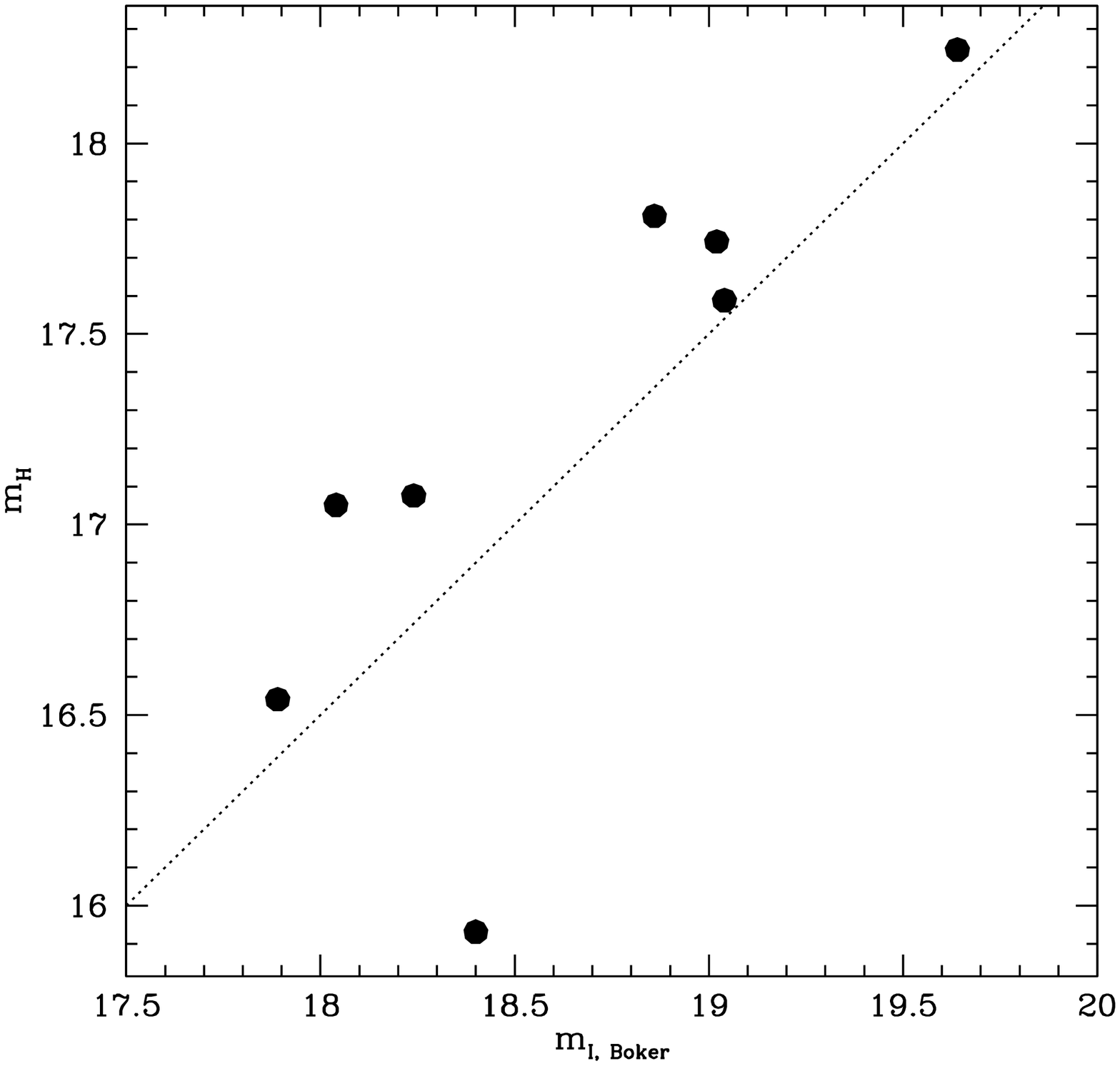}}
\end{center} 
\caption{Apparent magnitude of the inner
component against those published by \citet{boker}. The solid line overplotted indicates the $I = H + 1.5$ line. 
The clear outlier below the solid line represents NGC\,1042.}
\label{fig16ch4}
\end{figure}
\end{center}

In recent years, several studies have shown that many spiral galaxies host in their innermost regions an extra component,
distinguishable in the images and/or  photometric profiles as an excess light above the (exponential) disk plus (S\'ersic)
bulge. Nuclei are common both in early- and late-  type galaxies. \citet{phillips} found unresolved bright  nuclei in six
out of 10 late-type spirals; \citet{matteus} noticed a compact star cluster in the centres of 10 very late-type spirals, 
within a program studying 49 objects; out of a sample of 77 Scd-Sm galaxies, \citet{boker}  detected a distinct component
in 59 cases. In some of the late-type spirals, these clusters are the only prominent source within a few  kiloparsecs from
the centre. In earlier-types their detection is complicated by the presence of a luminous and extended  bulge, but
\citet{marcella98b} showed that nuclear clusters are present also there. As already mentioned, \citet{balcellsnuc} found a
central light excess  in 90 percent of a sample of S0-Sbc galaxies using NICMOS imaging, detecting both unresolved and
extended  sources; according to the authors, the extended inner components are geometrically flat systems and could be
inner disks, rings or bars;  the unresolved ones are most likely star clusters and are found in more than one third of the
galaxies with inner component.\\ 
\indent In our sample, two galaxies out of 18 (NGC\,864 and NGC\,4102) present a central
depression and the remaining 16 show a central light excess above the best-fit  to the bulge; one of these (NGC\,5678) is
an ambiguous case, possibly because of the very extended dust lanes affecting  this galaxy, but we included it in the
`central component' group. The numbers are very similar to those of \citet{balcellsnuc}. Out of these 16,  eight belong to
the sample studied by B\"oker et al. (2002): all the Scd and Sd galaxies in our  sample. The authors detect nuclear star
clusters  in all of them and publish an estimate of their size and luminosity in the $I-$ band.\\
\begin{table} \centering
\small{\begin{tabular}{@{}l | l l l }
\hline \hline
NGC & m$_{inner}$& M$_{inner}$& $V-H$ \\
(1) & (2) & (3) & (4) \\
\hline
     488 &        14.10 &       -18.44     &3.196\\
     628 &        16.27 &       -13.68     &2.644\\
     772 &        14.30 &       -18.46     &     \\
     864 &        16.86 &        -13.02    &2.863\\
    1042$^*$ &        15.93 &       -15.35 &     \\
    2805$^*$ &        17.74 &       -14.51 &     \\
    2964 &        15.45 &       -16.12     &3.496\\
    3346$^*$ &        18.25 &       -13.14 &     \\
    3423$^*$ &        17.59 &       -13.25 &     \\
3949 &  	 17.10 &       -13.71	  &2.695\\
4030 &  	 16.19 &       -15.42	  &3.027\\
4102 &  	 15.07 &       -15.55	  &3.499\\
4254 &  	 15.60 &       -15.84	  &3.271\\
4487$^*$ &	 16.54 &       -14.29 &\\
4775$^*$ &	  17.05 &	-14.71 &\\
5585$^*$ &	  17.08 &	-12.50 &2.281\\
5668$^*$ &	  17.81 &	-14.08 &\\
5678 &  	 15.77 &       -16.71	  &3.657\\
\hline
\end{tabular}}
\caption[\small Apparent and absolute magnitudes of the inner components]{\small Column (1): NGC identifier of the galaxies with measurable inner component; 
columns (2): apparent magnitude m$_{inner}$ of the inner component, in $H-$ band; columns (3): absolute magnitude M$_{inner}$ of the inner component, 
in $H-$ band. Columns (4): Central colours in apertures of 0.227$''$ diameter (0.500$''$ for NGC 5585) for the
galaxies with both HST F606W and F160W data. The asterisks marks the galaxies in common with the sample of \citet{boker}.}
\label{tab5ch4}\end{table} 
\indent For all the 16 galaxies, we built model  images of the bulges on the basis of the fit parameters, and subtracted
them from the HST `bulge images' (NICMOS or WFPC2-F814W depending on the object). The images that we obtained contain by
construction the residuals from the global (exponential) disk + (S\'ersic) bulge fit to the galaxy:  dust features, spiral
structure, bars, and the inner component in which we are interested. We estimated the HWHM of the central component using
the {\small{IRAF} } {\small{rimexam}} task, resulting in values ranging between $\approx$ 0.07 and 0.23\arcsec. 
We then
calculated the flux enclosed in a circle of radius twice the HWHM  and converted it to a magnitude scale ($H-$ band).
For NGC 4102 and NGC 864 we calculated the flux by measuring all the light in the inner aperture of diameter 0.227$''$.
The computed magnitudes (both apparent and absolute) are
listed in Table  \ref{tab5ch4}, where an asterisk marks the galaxies in common with the sample analysed by \citet{boker}.
For all galaxies for which both HST F606W and F160W data are available, we also calculated the central $V-H$ 
colour in an aperture of diameter 0.227$''$ (0.500$''$ for NGC 5585), and also tabulated these in Table  \ref{tab5ch4}.
Fig. \ref{fig16ch4} shows the comparison between the apparent  magnitudes that we calculated and those published by
\citet{boker}, for the inner component of the galaxies in common. The magnitudes of B\"oker et al.  refer to the $I -$
band, while our data are $H -$ band data (see Table \ref{tab6ch4}). The only galaxy in 
common with B\"oker et al. (2002) is NGC 5585, which is representative for the galaxies in this group, and clearly shows
much less extinction in its $V-H$ color map. We also see that the $V-H$ colors of the nuclei are similar within 0.2 mag to
the colors in the inner aperture with diameter 2.4$''$ ($\Delta((V-H)_c - (V-H)_{nuc}) ~=~ -0.007 \pm 0.199 ~{\rm mag}$). 
Even though only about 10\% of the light in the larger aperture
comes from the nuclear cluster, it shows that the stellar populations of the nuclear cluster are not decoupled from those
in the rest of the galaxy, as is e.g. the case in globular clusters in galaxies. 

Walcher et al. (2006) took high resolution spectra for a number of the Scd, Sd 
and Sm galaxies of B\"oker et al. (2002) and analysed them using stellar
population synthesis, allowing the extinction to be fitted as a free parameter.
They find an average extinction in $I$ of $\sim$ 0.4 mag, which is very
comparable to what we find (A$_V$ = 0.68, with a scatter of 0.55): When using the
Galactic extinction law (Rieke \& Lebofsky 1985) the extinction in $V$ is 1.64
$\times$  the extinction in $I_C$, or 0.66 mag. Given the fact that the galaxies
of Walcher et al. are later type than ours, we expect less extinction than in
our sample. However, large variations probably exist from galaxy to galaxy, and also the amount 
of extinction that one obtains varies with method. We can only conclude that extinction is important
in late-type galaxies. Note that NGC 1042, in common with our paper, has by far the largest extinction
(A$_I$ = 1.35 mag), consistent with Figure \ref{fig16ch4}. 

\section[Summary and conclusions]{SUMMARY AND CONCLUSIONS}\label{conclusionsecch4}
In this Paper we performed a photometric bulge-disk decomposition for our sample
of 18 late-type spiral galaxies.  We used NIR ($H-$ band) archive images (DSS,
2MASS and HST) from which we retrieved photometric profiles covering the whole
disk and having a high spatial resolution in the inner parts. We fitted an
exponential disk and a S\'ersic bulge, using a method that allows the two
components to have different intrinsic shapes, but without going to a full
two-dimensional fitting of the light distribution; we studied in detail the
correlations between the fit parameters, and we investigated the high resolution
$V-H$ colours in this sample, in the nuclear cluster and outside it. 
From the $V-H$ colour maps and the Mg~b maps of Ganda et al. (2007) we made 
extinction maps for 10 of the galaxies. We summarize below the main results.

\begin{itemize}
\item Late-type spirals can be fitted well by a model consisting of a S\'ersic bulge, an exponential disk, 
and a nuclear light access. Outside the central regions, the bulk of the galaxy's light is well
described by one exponential disk, although in some cases a double exponential
(inner + outer disk) would provide better fits, as tested by \citet{pohlen}.
In most cases bulges are rounder than their disks, but for the latest
types the opposite sometimes holds, possibly in relation with the presence of a
bar.
\item Bulges of late-type spiral galaxies follow the same relation between 
central velocity dispersion and bulge luminosity as bulges of early-type spirals.
This might indicate that the black hole - sigma relation also holds for late-type spiral
bulges. Late-type bulges have, however, lower surface brightness and are larger than early-type
bulges.
\item The structural parameters of our bulges agree statistically with the 
large compilation of Graham \& Worley (2008), showing a strong correlation between
surface brightness of bulge and disk, and a less strong relation between bulge effective
radius and disk scale length.
\item The star formation time scale inversely correlates with the S\'ersic index $n$,
implying that exponential bulges have long star formation time scales, in agreement with
the prediction from secular evolution models, that exponential bulges are disks forming
through secular evolution.
\item In 16 out of 18 galaxies the galaxy's profile presents an excess light
with respect to the fitted disk + bulge, which can be interpreted as an
additional tiny component, in many cases (in all the later types) a nuclear star
cluster. We give an estimate of the magnitude of these inner components.
\item Using $V-H$ colour maps from HST imaging and SAURON Mg~b maps, we have 
determined model-independent extinction maps for 10 of the 18 galaxies. We find that the central
A$_V$ ranges from 0 to 2 mag, with many galaxies being optically thick in the optical. The 
$V-H$ profiles show a lot of structure, mostly due to extinction, with much larger 
gradients than are generally displayed by early-type galaxies.
\item The colours of nuclear clusters are the same as the inner regions of their host galaxies,
with a scatter of 0.2 mag, indicating a similar composition and age as the stellar populations
just outside the center, unlike e.g. galactic globular clusters.
\end{itemize}

\section*{Acknowledgements}
We thank Alessandro Boselli and Luca Cortese for kindly making available
profiles useful in establishing the reliability of the methods used in this
Paper. We also thank Leslie Hunt, who very gently provided us her data in
tabular format, and Michael Pohlen, who provided us his profiles as well. We
kindly acknowledge Edo Noordermeer for very useful discussions and both him and
Alister Graham for providing FORTRAN codes used in the preparation of the
present Paper. Thanks to Isabel P\'erez, Gert Sikkema and Michele Cappellari for
helpful discussions and suggestions. KG acknowledges support for the Ubbo
Emmius PhD program of the University of Groningen. This project made use of the
HyperLeda and NED databases. Part of this work is based on data obtained from
the STSci Science Archive Facility. The Digitized Sky Surveys were produced at
the Space Telescope Science Institute under U.S. Government grant NAGW-2166. The
images of these surveys are based on photographic data obtained using the Oschin
Schmidt Telescope on Palomar Mountain and the UK Schmidt Telescope.

\appendix

\section[Internal consistency of the method - literature comparison]
{Internal consistency of the method - literature comparison}

\begin{figure*}\centering
{\includegraphics[width=0.8\textwidth]{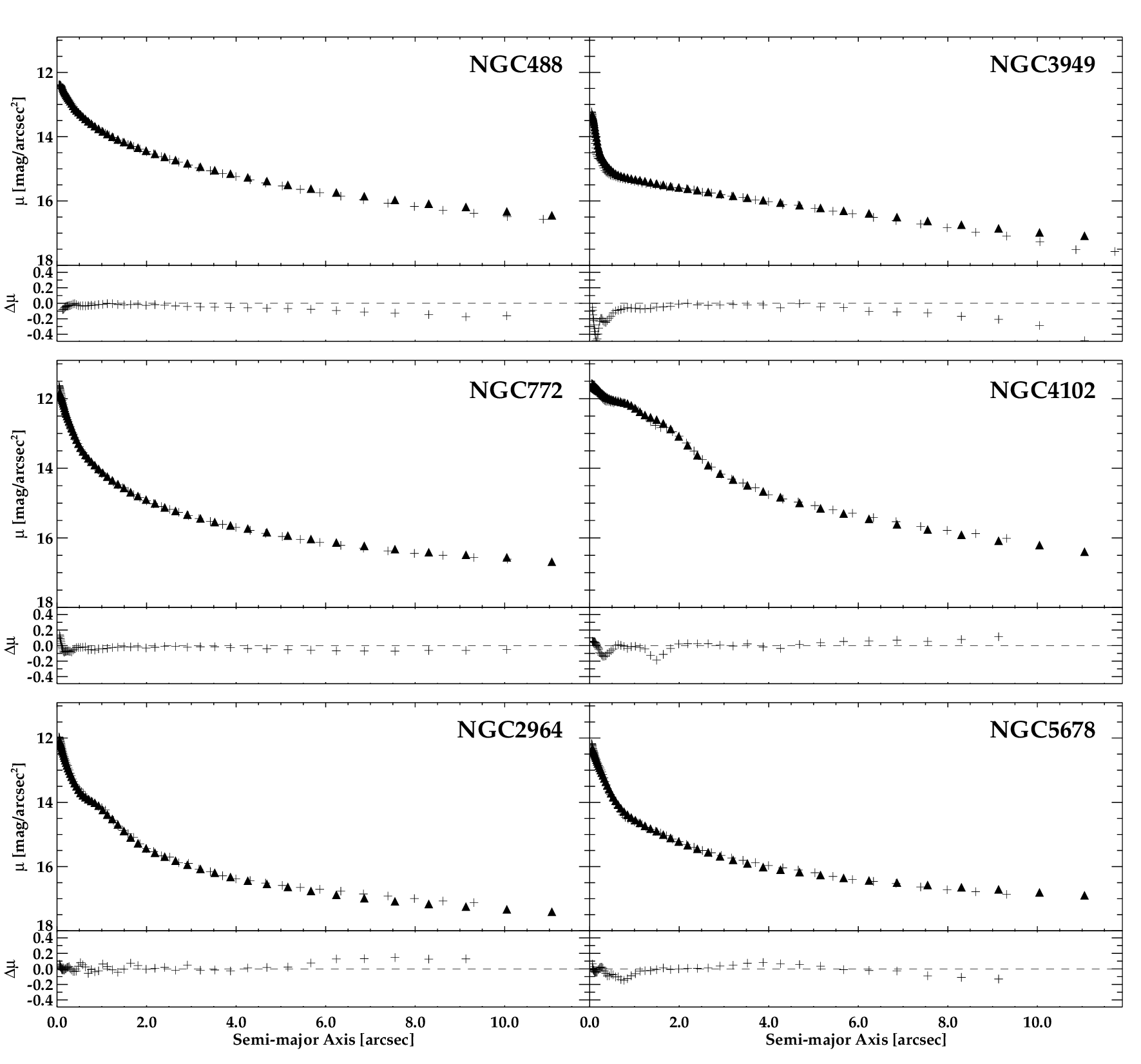}}
\caption{Comparison between our combined NICMOS + 2MASS profiles and the
profiles published by \citet{hunt} for the six galaxies in common, plotted on
the common radial range. In the left panels, the crosses represent our profiles,
and the filled triangles those kindly made available by Leslie Hunt, also
plotted on the common radial range. In the right panels, we plot the difference
between the two. The difference is relevant only in the innermost
0.1-0.3\arcsec, and might be due to small differences in the centering and/or in
the geometric parameters. In particular, NGC\,3949 represents the case with the
largest difference in the innermost points, and it is also the case where the
difference between the ellipticity profile from Hunt and the mean value we
adopted is the largest.}
\label{fig4ch4}
\end{figure*}

\begin{figure*} \centering
{\includegraphics[width=0.8\textwidth]{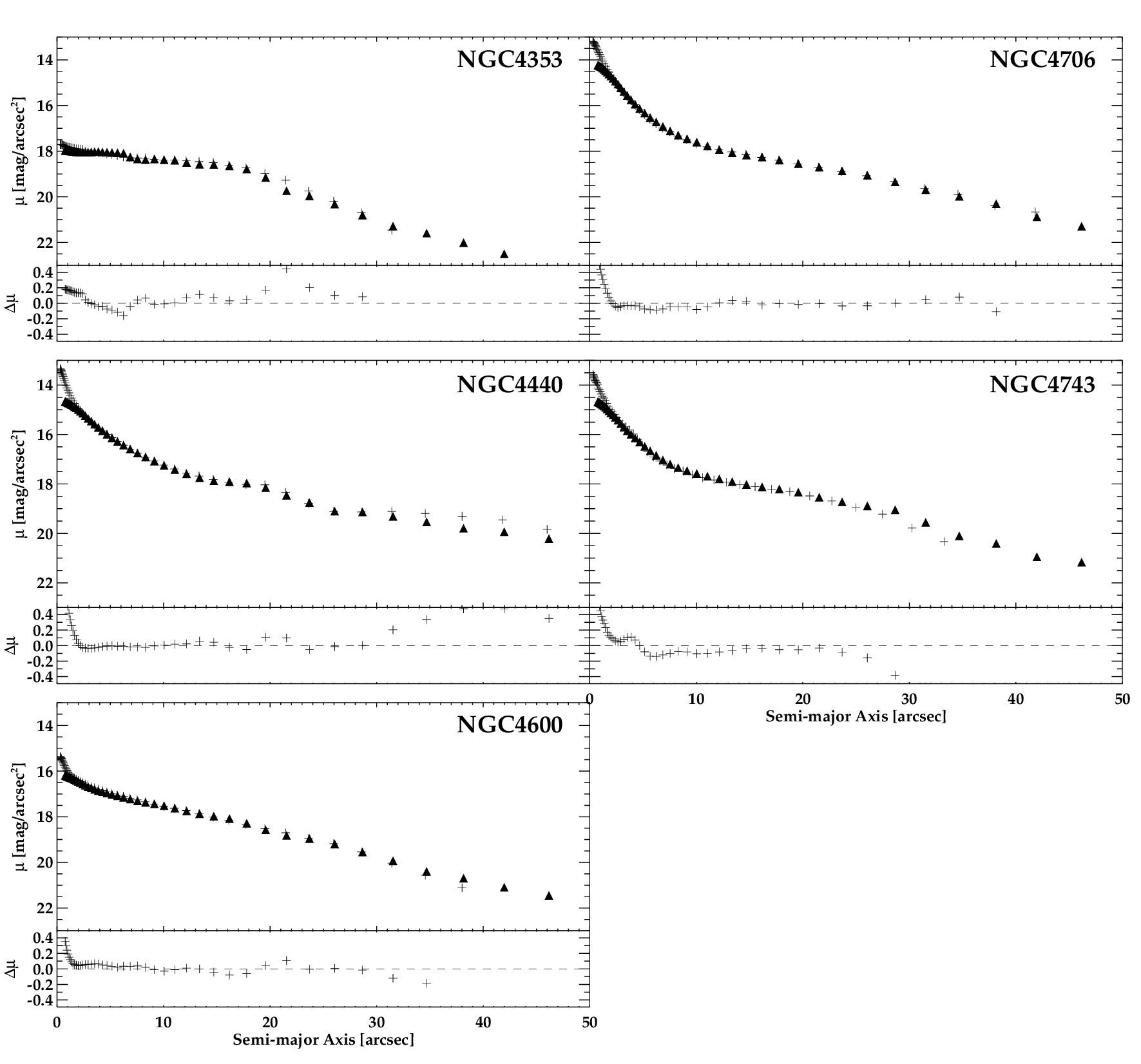}}
\caption{Comparison between the combined 2MASS-DSS profiles and the profiles
from \citet{boselli} for five galaxies taken from Gavazzi et al. (2001, see text
for a more detailed description). In the left panels, the crosses represent our
profiles, the filled triangles those from \citet{boselli}; the vertical scale is
in absolute $H$ magnitudes arcsec$^{-2}$. In the right panels, we show on the
common radial range the difference in magnitudes arcsec$^{-2}$ between the
profiles from \citet{boselli} and ours. In most cases, the differences are
significant (more than 0.1 mag arcsec$^{-2}$) only in the innermost 2-3\arcsec,
where our 2MASS profiles are affected by seeing smearing much more than the ones
of \citet{boselli}, and at radii larger than 25-30\arcsec, corresponding to the
external limit of the profiles from \citet{boselli}. This proves that the
agreement is good.}
\label{fig5ch4}
\end{figure*}

In order to ensure the reliability of the method applied, we searched the
literature for published $H-$ band profiles. For the late-type spiral galaxies
of our sample extremely little is available. One of the few existing references
is the work of \citet{hunt}, who present NICMOS photometry for 250 active and
normal galaxies, among which six are in common with our sample:
NGC\,488,\,772,\,2964,\,3949,\,4102,\,5678. In almost all cases they used the
same images as we did. The authors were so kind to provide us in tabular format
the profiles for those galaxies, that they extracted keeping the centre of the
isophotes fixed, but allowing the geometric parameters to vary over the whole
radial range. In order to match their approach, we extracted again photometric
profiles from the 2MASS and NICMOS images (DSS profiles are here redundant,
since the literature profiles are available only in the inner $\approx$
10\arcsec only; therefore we did not use them) along the average position angle
and ellipticity of the profiles from \citet{hunt}. In Fig. \ref{fig4ch4} we show
for those six galaxies the comparison with \citet{hunt}: the profiles (left
panels) and the difference between them, plotted on the common radial range. The
agreement is satisfactory (see also figure caption). NICMOS profiles for five of
the six galaxies in common with \citet{hunt}
(NGC\,488,\,772,\,2964,\,3949,\,5678), again derived from the same images that
we used, are published also by \citet{seigaretal}; from a visual inspection of
their profiles, we noticed that there is a significant offset with ours and
Hunt's profiles, which is probably due to the adoption of another magnitude
system (AB magnitude system), therefore we did not proceed to a more detailed
comparison.

As a reference for the ground-based $H-$ band photometry, we chose to use the
work of \citet{boselli}, who present $H-$ band observations and surface
brightness profile decomposition for 75 faint galaxies, mostly dwarf ellipticals
belonging to the Virgo cluster. They collected $H-$ band images at the ESO-NTT
telescope in La Silla (Chile) and at the TNG on La Palma (Spain), reduced and
calibrated the images and extracted brightness profiles fitting ellipses with
ellipse centres, ellipticities and position angles left as free parameters. This
corresponds to the first step in our procedure, as we described above. Since
there is no overlap between their sample and ours, we chose some galaxies out of
their paper (NGC\,4353, 4440, 4600, 4706 and 4743), retrieved the 2MASS and DSS
images for these galaxies and extracted brightness profiles with free geometric
parameters, subtracted an estimate for the sky background, combined them into a
single profile and converted the latter to an absolute magnitude scale in the
same way as done for our own galaxies -we only skipped the steps where we fix
the geometric parameters, for consistency with the approach used by
\citet{boselli}. Fig. \ref{fig5ch4} shows for the chosen galaxies the comparison
between our profiles and those published by \citet{boselli}, kindly made
available in tabular format by the authors. The figure shows that the agreement
is good, confirming the reliability of our method.

\section[Structural parameter relations]
{Structural parameter relations}
\label{strucpar}

In this appendix we present a large number of relations, between structural
parameters and luminosity of bulge, disk and total galaxy, and between
structural and stellar population parameters. A few of these have already been
discussed in Section 5. For the relations between structural
parameters and luminosity of bulge, disk and total galaxy we find that our late
type spiral galaxies generally behave in the same way as the compilation of
Graham \& Worley (2008). In Section 5.1 and 5.2 the most important correlations
were discussed. Here we discuss more relations, for completeness.

We start looking at the relations between the structural parameters of bulge
and disk and global galaxy properties such as morphological type and velocity
dispersion. We plot these parameters against each other in Fig. \ref{fig10ch4aa}
and
\ref{fig10ch4b}, for our sample with the compilation of GW always plotted as
well to guide the eye.
If we focus on our sample only, we can notice some trends: $n$ tends to decrease
going to later types, and both the 
effective surface brightness of the bulge and the central surface brightness 
of the disk decrease going to later types. These correlations agree with GW and
B07. Interesting to note is that for galaxies earlier than Sbc surface
brightness of bulge and disk is virtually independent of type. For later types
there is a strong correlation. There galaxies become smaller, and have lower
surface brightness of both bulge and disk, for increasing morphological type
(see also de Jong 1996,  paper III). We confirm that there is no strong
corelation between the size of bulges and disk and morphological type. Also, the
size ratio r$_e$/h is independent of type, and we do not find the 
mild decrease that \citet{lauren} claim. See GW for an extended discussion about
this ratio. The most important correlations with central celocity dispersion have been
discussed in Section 5.1. We find that the effective radius of the bulge does not
correlate with $\sigma$, nor does the scale length of the disk, or the ratio of bulge 
effective radius to disk scale length. There is a good correlation between bulge 
luminosity and $\sigma$ (see also B07) and between bulge to disk ratio and $\sigma$.

In Fig.~\ref{fig11ch4a} and \ref{fig11ch4b} we plot the structural parameters of
bulges and disks as a function of bulge, disk and total $H$-band luminosity.
We see that surface brightness of bulges and disks increases with increasing 
bulge or disk luminosity, consequence of the constancy of disk scale length 
and bulge effective radius. Since later type spiral galaxies tend to be fainter,
we find that the shape parameter $n$ increases with the luminosity of the bulge,
disk or the whole galaxy. The relation between the bulge magnitude M$_b$ and $n$
(top right panel in Fig. \ref{fig11ch4a}) is particularly tight: the correlation
coefficient $c$ is $\approx -0.7$, and it becomes even tighter ($c \approx
-0.92$) when removing the outlier point representing NGC\,864. As shown by
B07 in their Figure 6c, spiral galaxies follow the same trend
as the Virgo ellipticals from \citet{caon} in the plane (M$_{Bulge}$, $n$).
We find a marginally significant correlation between
bulge luminosity and effective radius, in the sense that brighter bulges are
also larger. When the whole compilation of GW is considered this correlation is
not found. In general, when the bulge luminosity increases, the S\'ersic index
$n$ decreases, and the surface brightness increases, with constant effective
radius. For late type spirals only, which all have a similar S\'ersic index
($n$=1), there is apparently a tendency that both $\mu_e$ and r$_e$ increase. 
The marginally significant correlation between disk luminosity and disk scale
length is probably caused by the two bright outliers. 
The B/D luminosity ratio increases as galaxy bulges become more luminous, while
it does not show strong trends with the disk magnitude M$_{Disk}$. 
In general, from
this work we infer that several correlations between structural properties of
spiral galaxies  are tighter versus M$_{Bulge}$ than versus M$_{Disk}$ (see
Table~\ref{tab3ch4}), in agreement with the conclusion of \citet{balcellsglobal}.
This is particularly intriguing, implying that (some of) the properties
are driven by the mass of the bulge. GW, who find the same strong relation, 
also comment that these new decompositions, with S\'ersic rather than de
Vaucouleurs bulge, show that very few spirals have B/T $>$ 1/3. In a long
discussion GW confirm this conclusion, and point out that the B/D ratios in
cosmological galaxy simulations are on the average much larger, showing that
the models, or probably the recipes for supernova feedback, need to be improved
considerably.

Fig. \ref{fig12ch4} shows the S\'ersic parameter $n$, the bulge
effective surface brightness and radius, the disk central surface brightness and
scale length, the $r_{e}$/h ratio and the B/D luminosity ratio as a function of 
H$\beta$, $\Delta$Mg{\textit{b}$'$}, age and star formation time-scale $\tau$. 
The stellar
population parameters refer to a central aperture of radius 1\farcs5.
$\Delta$Mg{\textit{b}$'$} is the difference in magnitudes between the observed
Mg{\textit{b}} index and the prediction, based on the observed velocity
dispersion, given by the Mg$_{2} - \sigma$ relation published by
\citet{Jorgensen}, which is now believed to hold for old spheroids (see
discussion in \citet{reynier11});
$\Delta$Mg{\textit{b}$'$} is therefore a measure of the importance of young
populations in the galaxy. For the sake of clarity, we remind the reader that 
$\Delta \mathrm{Mg{\textit{b}'}}$ can be derived from $Mg~b$ and $\sigma$ in the
following way,
adopting the Mg$_{2} - \sigma$ relation from Jorgensen, expressed as in Eq. 3 in
the paper of \citet{ganda2}, the definition of Mg{\textit{b}$'$} inferred from
Eq. 2 in the same paper and the conversion between Mg$_{2}$ and
Mg{\textit{b}$'$} given in Eq. 4 there:

\begin{equation}
\Delta \mathrm{Mg{\textit{b}'}} = -2.5 \times \log(1 - \mathrm{Mg{\textit{b}}}/32.5) - 0.096 \times \log(\sigma) + 0.062. 
\end{equation}

Going back to Fig. \ref{fig12ch4} and to the description of the quantities 
involved here, the age is the Single Stellar Population (SSP)-equivalent age 
inferred
by comparison of observed line-strength indices with models and $\tau$ is the
$e$-folding time-scale for star formation, assuming a galaxy age of 10 Gyr and
an exponentially declining star formation rate with time. For details we refer
the reader to Sections 5.3, 6.1 and 6.3, in \citet{ganda2}. 
We can observe
some trends, a few of which significant ($\vert c \vert \geqslant 0.5$), in
those cases we have also plotted the correlation as a dashed line: 
when H$\beta$ increases, i.e. when the stellar population becomes younger, 
$n$ decreases, both $\mu_{0,Disk}$ and $\mu_{e,Bulge}$ become
fainter, and the disk scale length h and the bulge effective radius $r_{e}$
decrease. 
There are also indications that both $r_{e}$ and h become smaller with
$\Delta$Mg{\textit{b}$'$} becoming more negative, which means that galaxies
farther away from the Mg$_{2} - \sigma$ relation of old ellipticals, i.e.
younger galaxies, have both
smaller bulges and smaller disks. 
No relevant trend is found with age: possibly
a confirmation of the fact that, in general, SSP-equivalent ages and
metallicities are not a good approximation for spiral galaxies, as discussed by
\citet{reynier11} and \citet{ganda2}. A more relevant parameter for our galaxies
seems to be $\tau$, in the framework of a continuous star formation: as $\tau$
increases (and the star formation history is less and less well described by a
single burst), $n$ becomes smaller, $\mu_{0,d}$ and $\mu_{e,b}$ become fainter
and $r_{e}$ and h become smaller. No significant trend is observed involving the
ratio $r_{e}/$h ratio and the B/D luminosity ratio. Globally, Fig.
\ref{fig12ch4} seem to corroborate a scenario where galaxies with a quiescent,
almost constant star formation, hosting composite stellar populations, with (at
least) some young stars, have almost-exponential bulges and small and faint
bulges and disks, while galaxies that formed stars in a single burst long ago
(more SSP-like) tend to have higher $n$ and bigger and brighter bulges and
disks. A similar scenario, in which the character of the star formation history
is replaced by the galaxy age, is suggested also by \citet{huntbd}, as we
mentioned earlier: exponential bulges would be young products of secular
evolution; as the galaxy ages, the bulges would grow and their $n$ increase.

\renewcommand{\thefigure}{\ref{strucpar}\arabic{figure}.\alph{subfigure}}
\setcounter{subfigure}{1}
\begin{figure*}\begin{center}
{\includegraphics[width=10truecm]{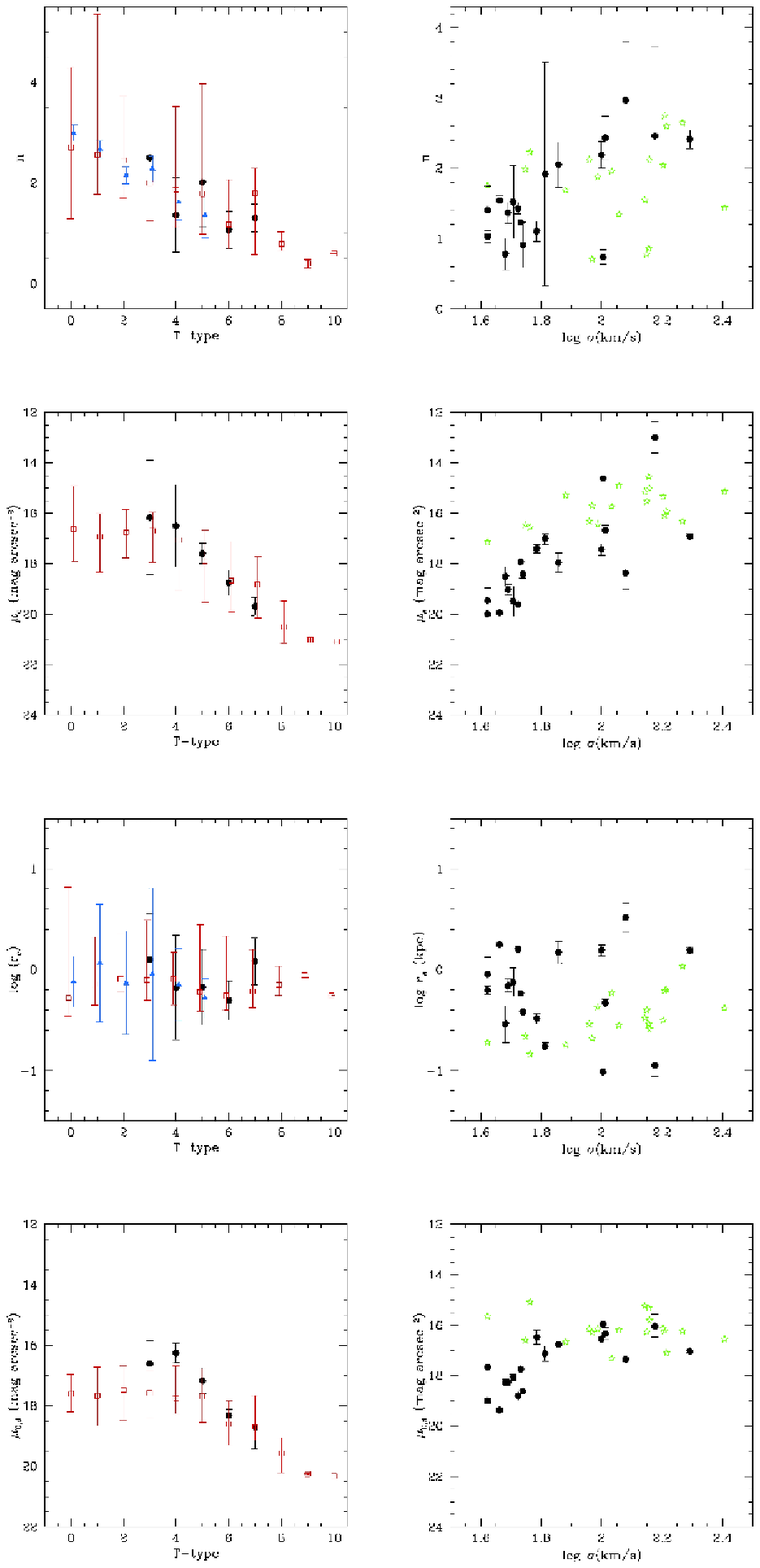}}
\caption{Bulge and disk structural parameters as a function of 
morphological T-type and central stellar velocity dispersion
$\sigma$. The black filled circles represent our sample; blue triangles
represent the sample of Fisher \& Drory (2008). The red open squares
represent the literature compilation of Graham \& Worley (2008). 
{\textit{Left panels}}: from top to bottom we plot $n$, $\mu_{e,Bulge}$, 
$r_e$ and $\mu_{0,Disk}$; for
each sample and each value of T-type, we plot the average value of the quantity of
interest for all the galaxies of that type in that sample, with an errorbar
indicating the standard deviation around the mean. {\textit{Right panels}}: we
plot the same quantities against log($\sigma$), measured in km s$^{-1}$. 
The velocity
dispersions for our sample are taken from Table 1 in \citet{ganda}. As for the
morphological T-type, we have artificially put all the galaxies earlier than T =
0 in the sample of \citealt{balcellsglobal}
to T = 0.}
\label{fig10ch4aa}
\end{center}\end{figure*}

\addtocounter{figure}{-1}
\addtocounter{subfigure}{1}
\begin{figure*}\begin{center}
{\includegraphics[width=10.5truecm]{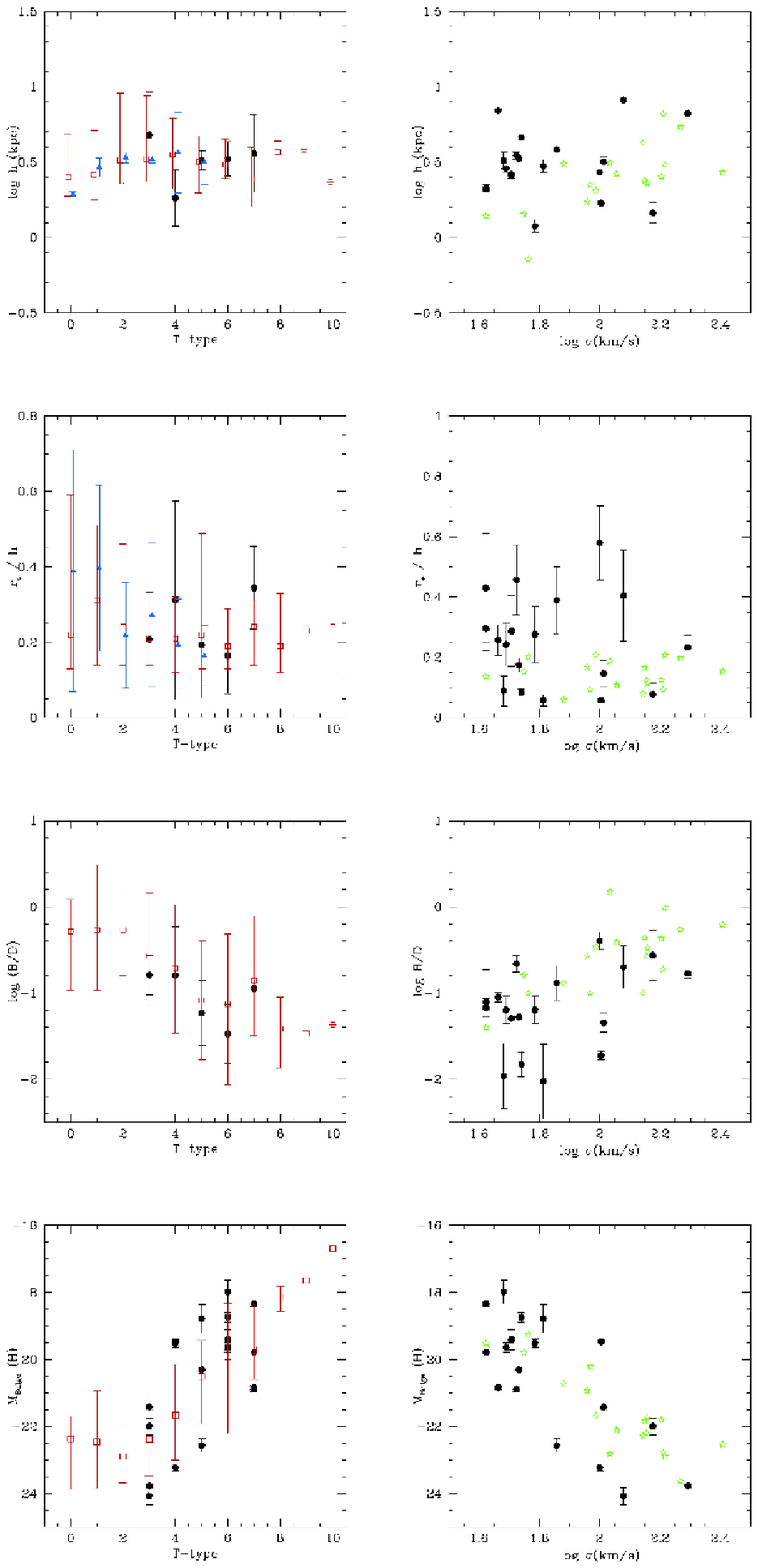}}
\caption{Bulge and disk structural parameters as a function of 
morphological T-type and central stellar velocity dispersion
$\sigma$ (see Fig. \ref{fig10ch4aa}).
{\textit{Left panels}}: from top to bottom we plot disk scale length, 
bulge effective radius to disk scale length ratio, $H$-band
bulge-to-disk ratio  and bulge luminosity against T; {\textit{right panels}}: 
same quantities plotted against log($\sigma$)in km s$^{-1}$.}
\label{fig10ch4b}
\end{center}
\end{figure*}

\renewcommand{\thefigure}{\ref{strucpar}\arabic{figure}.\alph{subfigure}}
\setcounter{subfigure}{1}
\begin{figure*} \begin{center}
{\includegraphics[width={0.9\textwidth}]{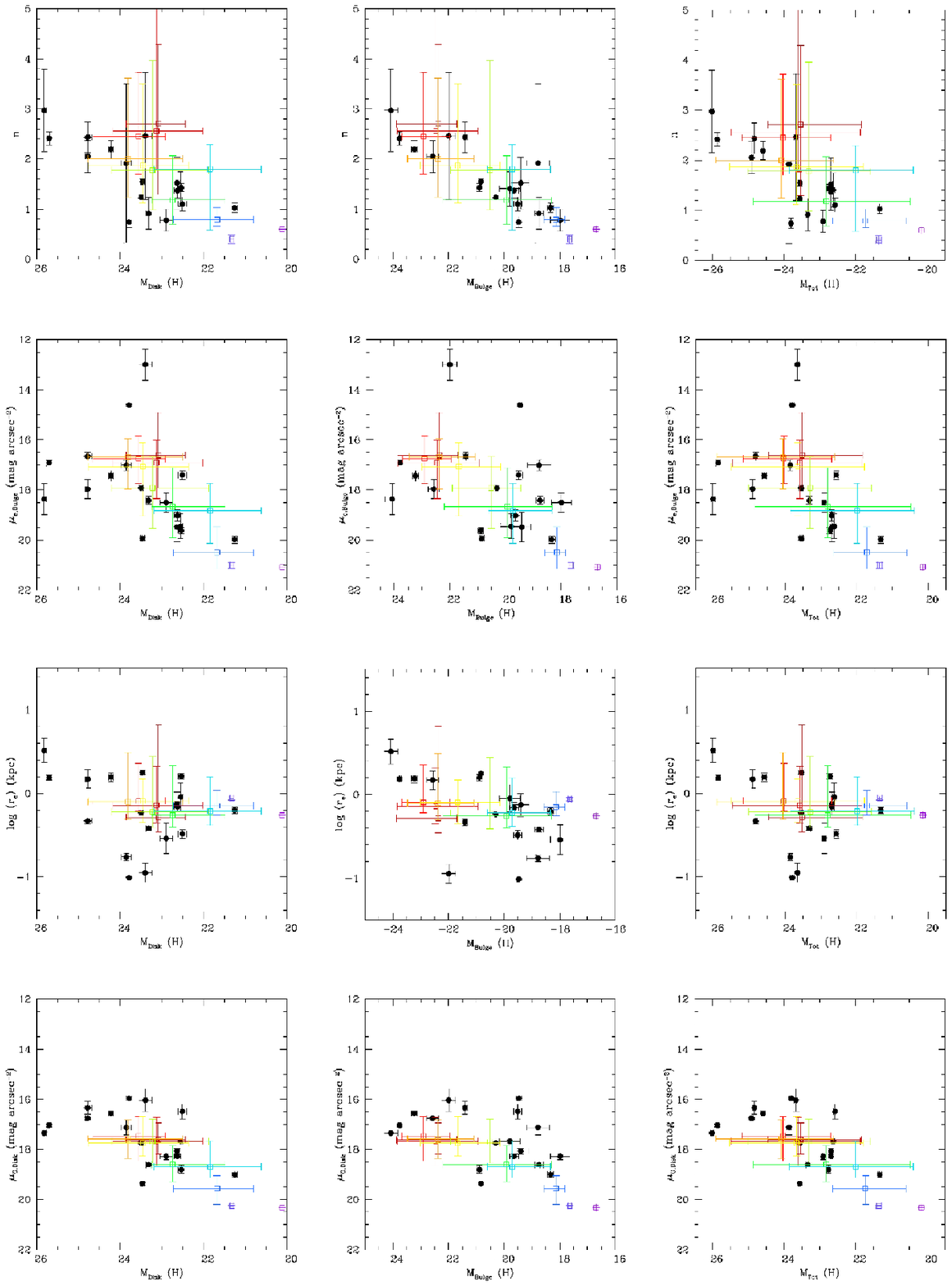}}
\caption{Correlations between structural parameters of bulge and disk with 
the luminosity of the disk (left), bulge (middle) and the whole galaxy (right). 
For symbols see Fig. \ref{fig10ch4aa}.
From top to bottom we plot correlations of S\'ersic index $n$, bulge effective surface
brightness, bulge effective radius and central disk surface brightness. The sample of 
Graham \& Worley has been plotted in different colours, ranging from red for the earliest
type galaxies to blue for the latest types.
}
\label{fig11ch4a}
\end{center}
\end{figure*}

\addtocounter{figure}{-1}
\addtocounter{subfigure}{1}
\begin{figure*} \begin{center}
{\includegraphics[width=\textwidth]{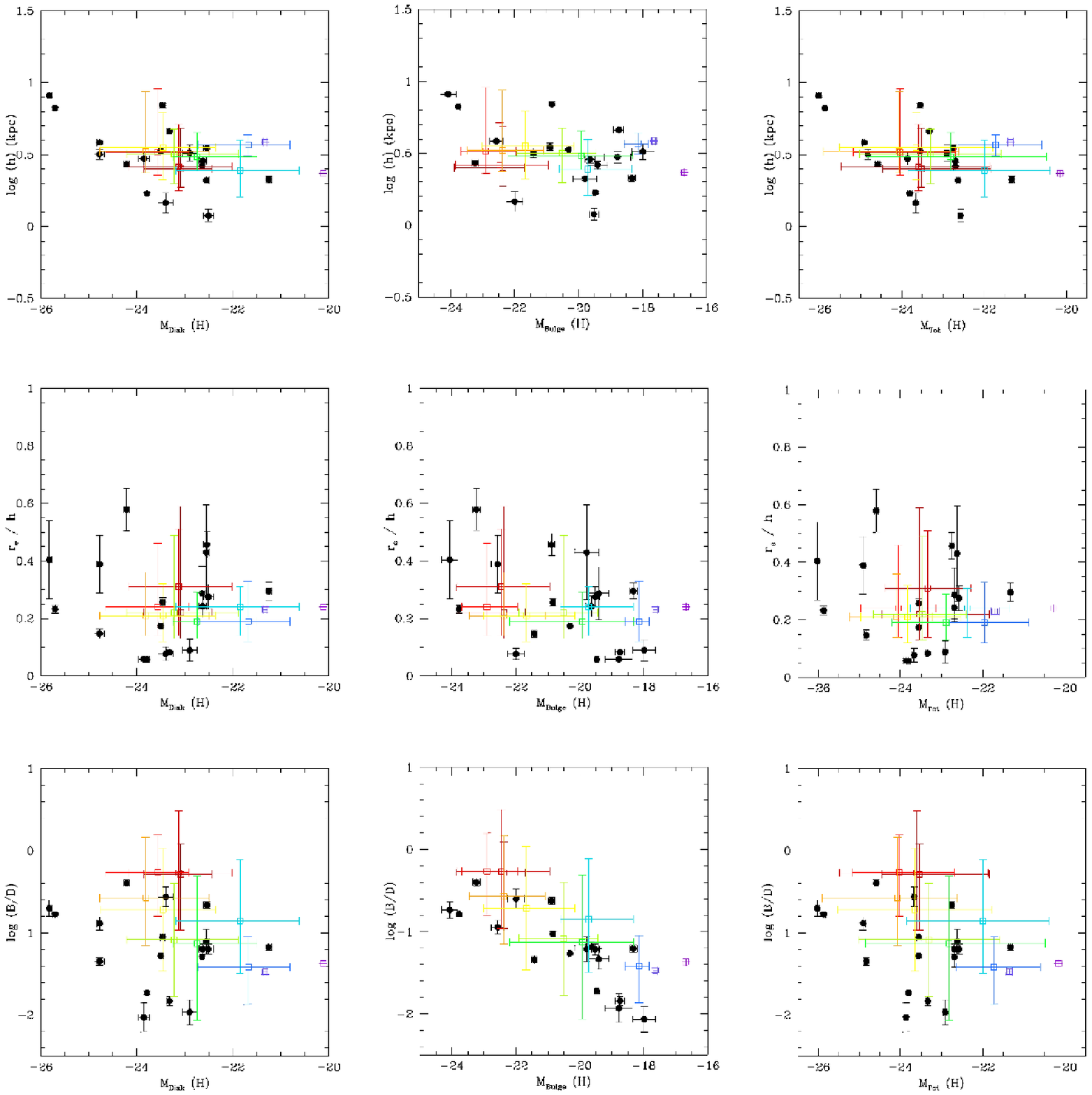}}
\caption{Correlations between structural parameters of bulge and disk with 
the luminosity of the disk (left), bulge (middle) and the whole galaxy (right).
{\textit{Left panels}}: from top to bottom we plot correlations with 
disk scale length, 
ratio of bulge effective radius to disk scale length,
and $H$-band bulge-to-disk ratio. The sample of 
Graham \& Worley has been plotted in different colours, ranging from red for the earliest
type galaxies to blue for the latest types.}
\label{fig11ch4b}
\end{center}
\end{figure*}

\renewcommand{\thefigure}{\ref{strucpar}\arabic{figure}}
\begin{figure*}\begin{center}
{\includegraphics[width=\textwidth]{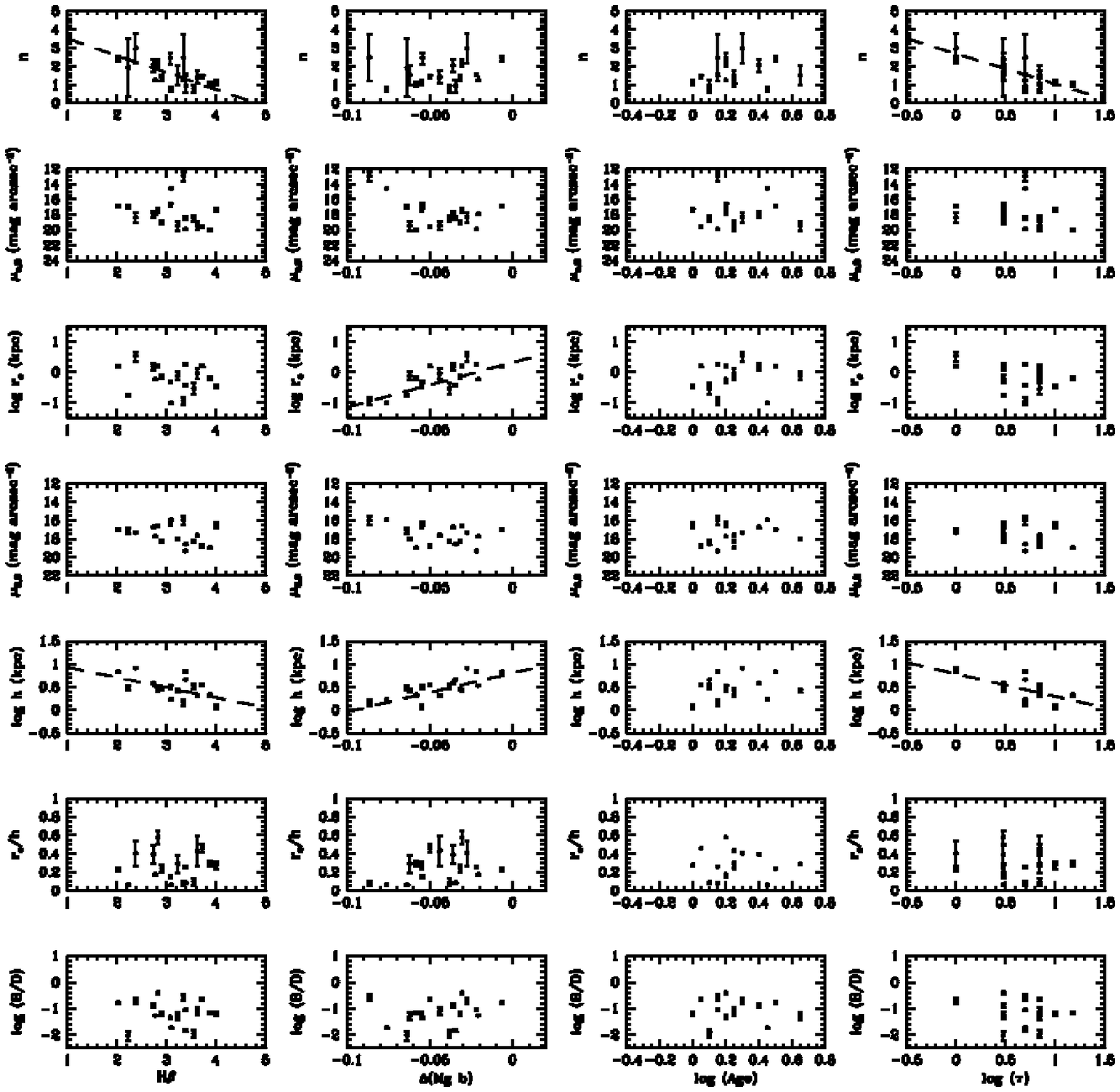}}
\caption{Correlations between the structural parameters $n$, $\mu_{e,b}$,
$r_{e}$, $\mu_{0,Disk}$, h, $r_{e}$/h and the B/D luminosity ratio against the
population parameters H$\beta$ (in \AA), $\Delta$Mg{\textit{b}}$'$ (in
magnitudes), age and $\tau$ (in Gyr). 
The dashed lines overplotted in some of the panels represent a linear
fit.}
\label{fig12ch4}
\end{center}\end{figure*}

%%%%%%%%%%%%%%%%%%%%%%%%%%%%%%%%%%%%

\makeatletter
% define \thebiblio (same as thebibliography, but
% without the section heading)
\def\thebiblio#1{%
 \list{}{\usecounter{dummy}%
         \labelwidth\z@
         \leftmargin 1.5em
         \itemsep \z@
         \itemindent-\leftmargin}
 \reset@font\small
 \parindent\z@
 \parskip\z@ plus .1pt\relax
 \def\newblock{\hskip .11em plus .33em minus .07em}
 \sloppy\clubpenalty4000\widowpenalty4000
 \sfcode`\.=1000\relax
}
\let\endthebiblio=\endlist
\makeatother

%%%%%%%%%%%%%%%%%%%%%%%%%%%%%%%%%%%%

% \bsp % ``This paper has been produced using the ...''

\label{lastpage}

\end{document}